\documentclass{jfm}
\usepackage{graphicx}
\usepackage{dcolumn}
\usepackage{color}
\usepackage{placeins}
\usepackage{array}
\usepackage{subfigure}
\usepackage{natbib}
\usepackage{amssymb,amsmath}
\usepackage{epstopdf}

\newcommand{\tpr}{\text{Pr}}
\newcommand{\tra}{\text{Ra}}
\newcommand{\tnu}{\text{Nu}}
\newcommand{\tre}{\text{Re}}

\begin{document}

\title[Comparison between two and three dimensional Rayleigh-B\'enard convection]
{Comparison between two and three dimensional Rayleigh-B\'enard convection}

\author[Erwin P. van der Poel, Richard J.A.M. Stevens and Detlef Lohse]
{Erwin P. van der Poel$^1$, Richard J.A.M. Stevens$^{1,2}$ and Detlef Lohse$^1$}

\affiliation{
$^1$Department of Science and Technology and J.M. Burgers Center for Fluid Dynamics, University of Twente, P.O Box 217, 7500 AE Enschede, The Netherlands.\\
$^2$Department of Mechanical Engineering, Johns Hopkins University, Baltimore, Maryland 21218, USA.
}

\pubyear{}
\volume{}
\pagerange{}
\date{\today}

\maketitle

\begin{abstract}
Two dimensional (2D) and three dimensional (3D) Rayleigh-B\'enard convection is compared using results from direct numerical simulations and prior experiments. The explored phase diagrams for both cases are reviewed. The differences and similarities between 2D and 3D are studied using Nu(Ra) for $\tpr = 4.38$ and $\tpr = 0.7$ and Nu(Pr) for Ra up to $10^8$. In the Nu(Ra) scaling at higher Pr, 2D and 3D are very similar; differing only by a constant factor up to $\tra = 10^{10}$. In contrast, the difference is large at lower Pr, due to the strong roll state dependence of Nu in 2D. The behaviour of Nu(Pr) is similar in 2D and 3D at large Pr. However, it differs significantly around $\tpr = 1$. The Reynolds number values are consistently higher in 2D and additionally converge at large Pr. Finally, the thermal boundary layer profiles are compared in 2D and 3D.
\end{abstract}

\section{Introduction}
In Rayleigh-B\'enard (RB) a fluid in a closed sample is heated from below and cooled from above. This system is widely studied due to its conceptual simplicity and because of the many applications of turbulent heat transfer, such as in geophysics, astrophysics or process technology. The control parameters of RB convection in the Boussinesq approximation are the Rayleigh number $\tra = \beta g \Delta L^3/(\kappa\nu)$, the Prandtl number $\tpr = \nu /\kappa$ and the aspect-ratio $\Gamma = D/L$. Here, $L$ is the height of the sample and $D$ its width, $\beta$ is the thermal expansion coefficient, $g$ the gravitational acceleration, $\Delta$ the temperature difference between the bottom and the top of the sample, and $\nu$ and $\kappa$ the kinematic viscosity and the thermal diffusivity, respectively. 

The response of the system is commonly quantified by the heat transfer and the kinetic energy, which we indicate with the Nusselt number Nu and the Reynolds number Re based on the root-mean-square vertical velocity, respectively.

\begin{equation}
\tnu = \frac{\langle u_z \theta \rangle_A - \kappa \langle \partial_z \theta \rangle_A}{\kappa\Delta L^-1},
\end{equation}

where $\langle \cdot \rangle_A$ indicates the average over any horizontal plane (3D) or line (2D), and the Reynolds number Re is defined as

\begin{equation}
\tre = \frac{u_{3}^{RMS}L}{\nu},
\end{equation}

where $u_{3}^{RMS}$ is the root-mean-square of the vertical velocity (i.e. parallel to gravity).

Though all real-world applications of RB convection are three-dimensional (3D), two-dimensional (2D) simulations are used to better understand the physical mechanisms of 3D convection, as 2D simulations are substantially less CPU-intensive than 3D simulations. In addition, theoretical predictions for scalings in hard turbulent RB convection (\cite{cas89},\cite{rob79},\cite{shr90},\cite{gro00,gro01,gro02,gro04,gro11}) are based on 2D equations, namely the Prandtl equations for the boundary layer or use assumptions that apply to 2D as well as to 3D. This makes it an useful tool to validate theory, regardless of the similarity between 2D and 3D. Moreover, the quasi-2D character of the large scale circulation (LSC) in both 2D and 3D flows hints towards a large similarity, in particular for integral quantities, between 2D and 3D in RB turbulence, in contrast to unbounded turbulence where in 3D no such large scale structures emerge. 

Despite these similarities, there are significant differences between 2D and 3D convection. For example, the limited motion of the LSC in 2D increases the accumulation of energy in corner-rolls leading to large scale wind reversals (\cite{sug10}) and high sensitivity of global output parameters on the roll-state (\cite{poe11}). In 3D these phenomena are also observed by \cite{wei10}, however the additional degree of freedom of the LSC attenuates the global effects of these phenomena. Furthermore, the intrinsic inverse energy cascade (\cite{kra67b}) of 2D turbulence is fundamentally different from the forward cascade of 3D turbulence. The effect of this difference at smaller scales on global properties is unfortunately unknown. However, one can argue that for RB flows with a large scale roll the effect must be minor, as in 3D RB the self-amplifying local driving and global temperature gradient \cite{ahl09} result in a box-sized vortex, even without an inverse energy cascade. The large scale dynamics are similar in 2D and 3D and therefore the main difference in global output between both systems is expected to come from the small scale dynamics.

Previous work of \cite{sch04} on the validity of the 2D approach to 3D RB convection concluded that for small Pr, 2D numerics are no longer a valid representation of 3D convection, due to the increasing energy in the toroidal component of the velocity in 3D flows at lower Pr (\cite{bus78b}). The analysis was limited to a low $\tra =10^6$, which is in the laminar regime for most Pr, according to the coherence length criterion of \cite{sug07}. Furthermore, \cite{sch04} used stress-free velocity boundary conditions on the lateral walls. Although this decreases computational requirement due to the absence of sidewall boundary layers, it complicates comparison to experiments where the boundary conditions are exhaustively no-slip. Now, eight years later, we are able to study the similarities and differences between 2D and 3D RB convection with no-slip boundary conditions at much higher Ra in the turbulent regime.

We explain the numerical methods used and provide resolution checks alongside the results. We show parameter spaces containing a comprehensive overview of available data points from 2D and 3D RB studies. A qualitative review is made using flow field snapshots of 2D and 3D flows at different Pr, illustrating the different regimes in Pr space and their proposed effect on the 2D-3D similarity. Additionally, the thermal boundary layer profiles obtained in the 2D and 3D simulations are compared with the 'flat-plate' Pohlhausen profile.

\section{Explored parameter space}
\begin{figure}
\centering
\subfigure{\includegraphics[width=0.49\textwidth]{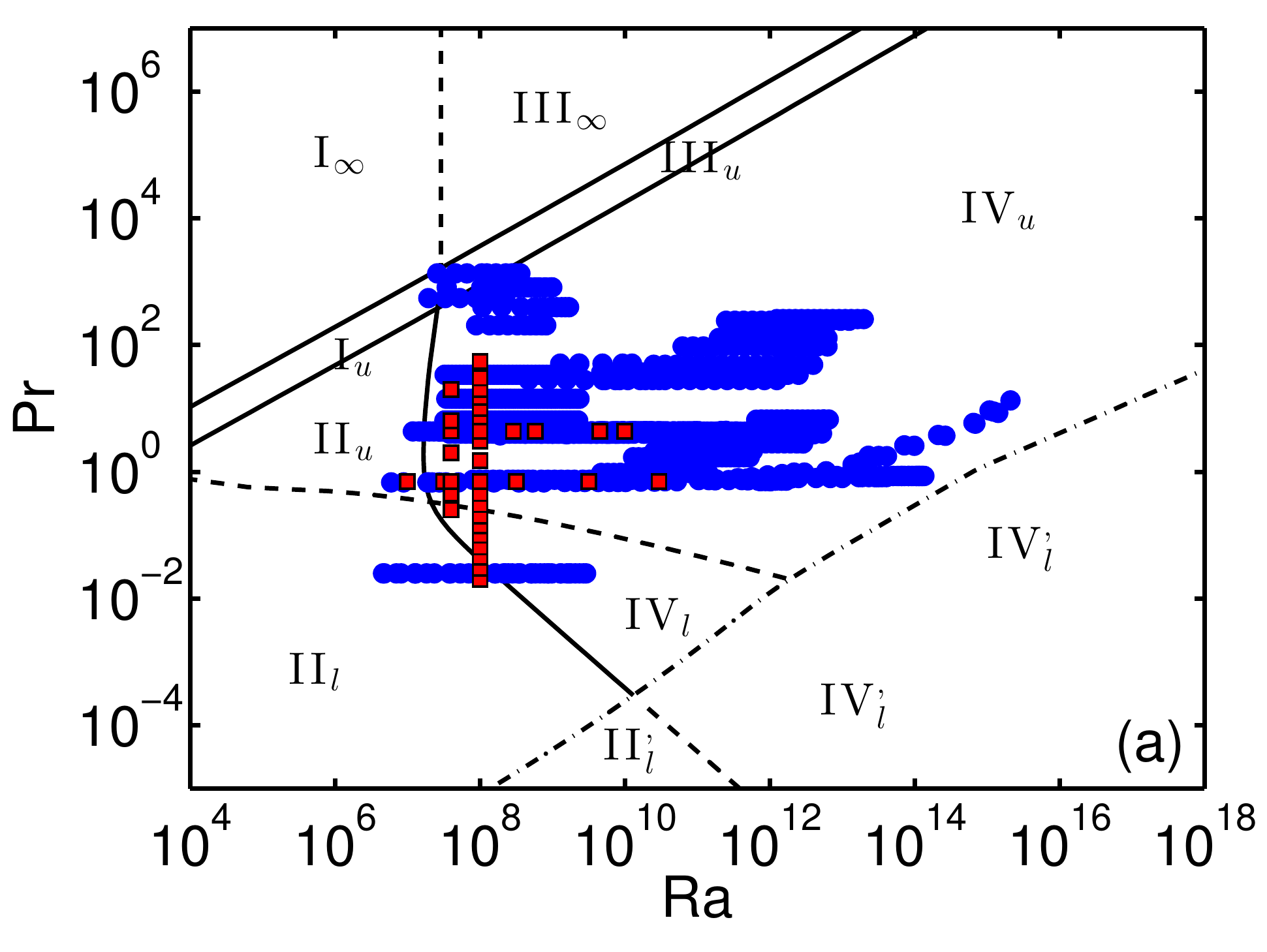}}
\subfigure{\includegraphics[width=0.49\textwidth]{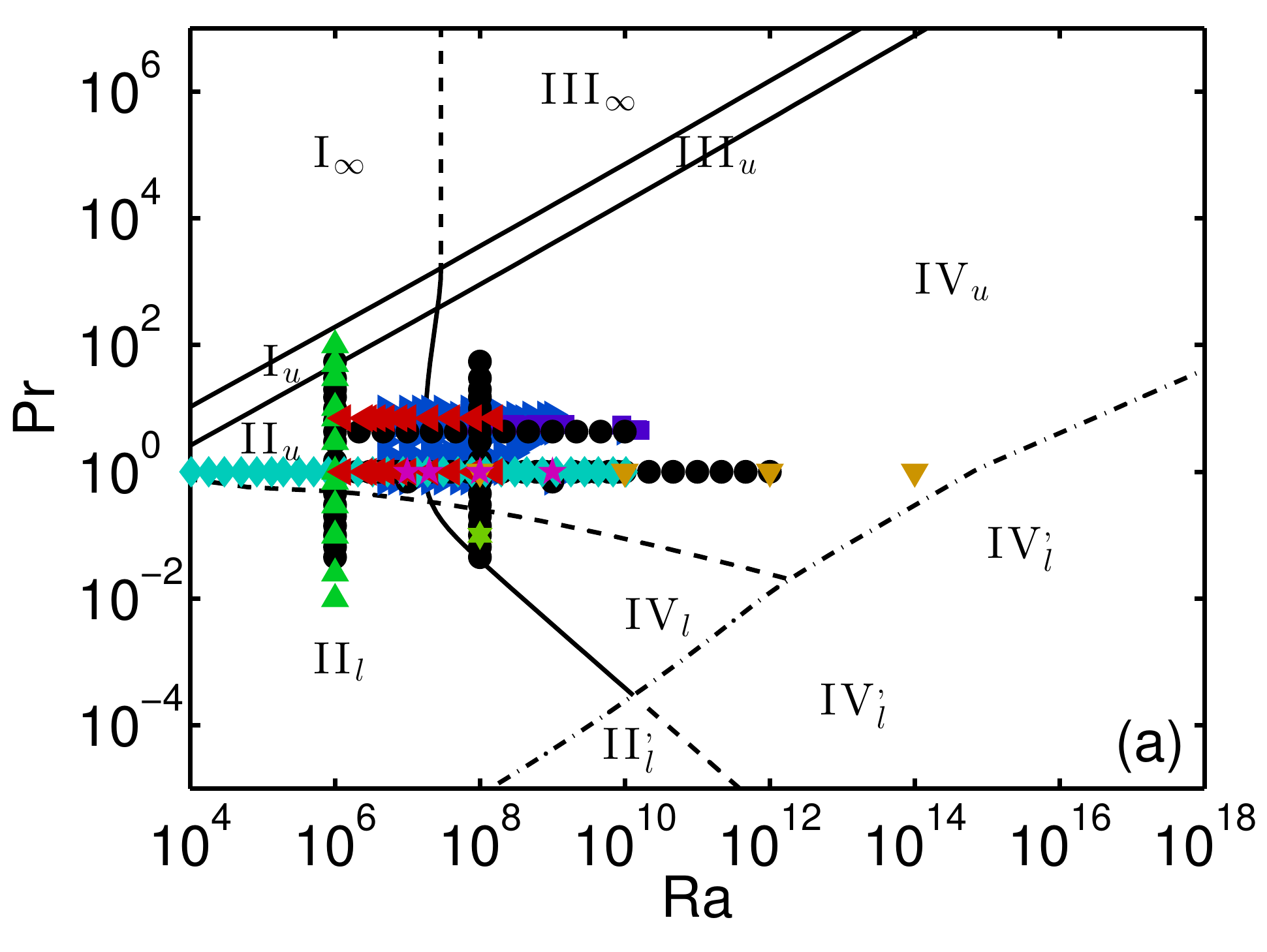}}
\subfigure{\includegraphics[width=0.90\textwidth]{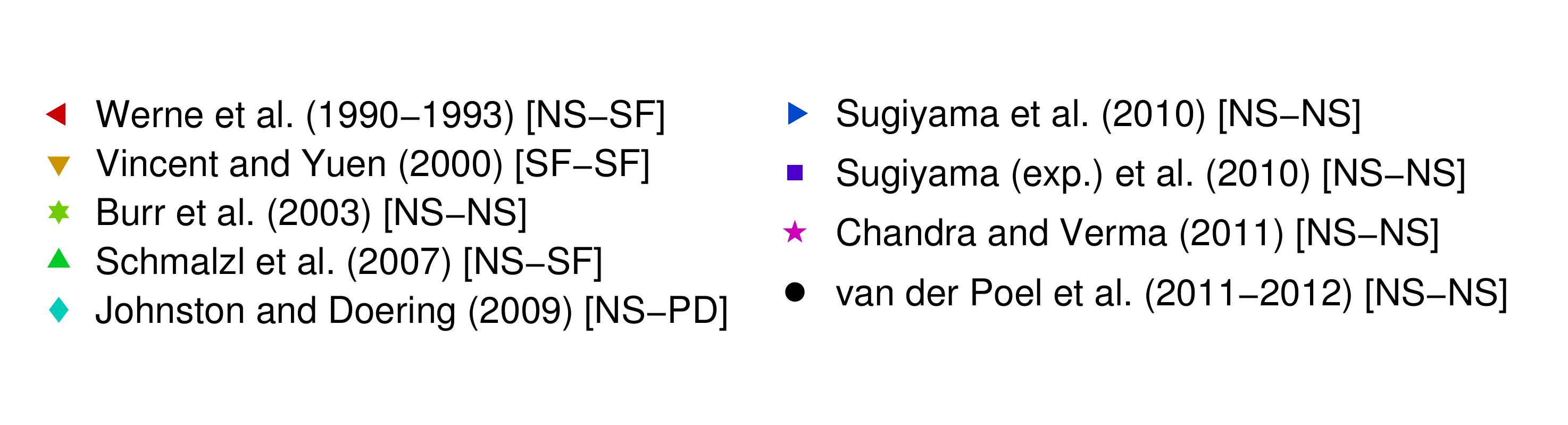}}
\caption{a) Phase diagram in Ra-Pr plane for three-dimensional studies. Here simulations and experiments are indicated by red squares and blue circles respectively. For more detail and references see \cite{ste13}. b) Phase diagram in Ra-Pr plane for two-dimensional studies. The symbol indicates the used boundary conditions; no-slip on all walls (circle), no-slip on horizontal plates and free-slip on sidewall (upward pointing triangle), free-slip on all walls (downward pointing triangle) and no-slip on horizontal plates and periodic sidewall (diamond). Note that the lines indicating the different GL regimes are taken from the 3D fit to assist the comparison of the two plots. The legend applies only to the 2D phase-diagram. The data points are taken from studies where Nu has been measured in all aspect-ratios and is from \cite{delu90}, \cite{wer91}, \cite{wer93}, \cite{vin00}, \cite{bur03}, \cite{sch04}, \cite{joh09}, \cite{sug10}, \cite{cha11}, \cite{poe11} and this study. Note that (exp.) in the legend signifies that this data is taken from quasi two-dimensional experiments. The lines in both plots are according to the Grossmann-Lohse theory (\cite{gro00,gro01,gro02,gro04}): The upper solid line means $\tre=\tre_c$; the lower nearly parallel solid line corresponds to $\epsilon_{u,BL}=\epsilon_{u,bulk}$; the curved solid line is $\epsilon_{\theta,BL}=\epsilon_{\theta,bulk}$; and along the long-dashed line $\lambda_u=\lambda_\theta$, i.e., $2a\tnu = \sqrt{\tre}$. The dotted line indicates where the laminar kinetic BL is expected to become turbulent, based on a critical shear Reynolds number $\tre_s^*=1014$ of the kinetic BL, with $a = 0.911$ (\cite{ste13})}
\label{fig:figure0}
\end{figure}

We first review the explored parameters space in both experiments and numerics. In figure \ref{fig:figure0} the phase diagram of 3D and 2D are displayed. The lines and numbers indicate the different regimes of the GL-theory based on a refit of the data (\cite{ste13}). Note that for the 2D plot the regimes resulting from the 3D fit are used. For 3D, data points where Nu has been measured or numerically calculated have been included for aspect ratio $\Gamma=1$ for no-slip velocity boundary conditions on all walls.

Unlike the 3D phase diagram in figure \ref{fig:figure0}, multiple velocity boundary conditions are included in the 2D phase diagram, which are indicated by the symbol; no-slip on all walls (circle), no-slip on horizontal plates and free-slip on sidewall (upward pointing triangle), free-slip on all walls (downward pointing triangle) and no-slip on horizontal plates and periodic sidewall (diamond). This increased variety in employed boundary conditions presumably originates from the lack of 2D experiments; there is less intention to mimic the no-slip experimental boundary conditions of experiments in 3D. In addition, the rectangular 2D geometry allows for more types of boundary conditions than the common cylindrical 3D setup as periodic sidewalls are not possible in this case. 

Comparing both phase diagrams, it becomes clear that the 3D parameter space is more explored due to the availability of experiments and the closer resemblance to convection in nature. The 2D phase diagram is, apart from one experimental series, fully composed of numerical data. The highest $\tra = 10^{14}$ is obtained by \cite{vin00} for a flow without velocity boundary layers. Evaluating the used grid resolution and their saturating Nu(Ra) data we think this simulation is underresolved and the heat flux was dominated by numerical diffusion. Discarding this point from the comparison and taking into account the fact that $\tra=2\cdot10^{12}$ is the highest Rayleigh number obtained in 3D, it is apparent that the exploration of the parameter space in 2D is open for a large improvement.
\section{Numerical simulations}
We numerically solve the three-dimensional Navier-Stokes equations within the Boussinesq approximation,
\begin{eqnarray}
 \frac{D\textbf{u}}{Dt} &=& - \nabla P + \left( \frac{\tpr}{\tra} \right)^{1/2} \nabla^2 \textbf{u} + \theta \textbf{$\widehat{z}$}, \\
 \frac{D\theta}{Dt} &=& \frac{1}{(\tpr\tra)^{1/2}}\nabla^2 \theta ,
\end{eqnarray}
with  $\nabla \cdot \textbf{u} = 0$.  Here \textbf{$\widehat{z}$} is the unit vector pointing in the opposite direction to gravity, $D/Dt = \partial_t + \textbf{u} \cdot \nabla $ the material derivative,  $\textbf{u}(\textbf{x},t)$ the velocity vector with no-slip boundary conditions at all walls, and $\theta$ the non-dimensional temperature, $0\leq \theta \leq 1$. The equations have been made non-dimensional by using the length $L$, the temperature $\Delta$, and the free-fall velocity $U=\sqrt{\beta g \Delta L}$. The 3D numerical scheme is described in detail in \cite{ver96} and \cite{ver99,ver03} and the 2D scheme in \cite{sug09}.

For this study we performed 3D simulations at $\tra=10^8$ and $ 0.02 \leq \tpr \leq 0.7$ in a $\Gamma=1$ sample and 2D simulations in a $\Gamma=1$ sample at $\tra=10^8$ with $ 0.065 \leq \tpr \leq 55$, and with $\tpr=4.38$ and $10^7 \leq \tra \leq 10^{11}$ to complement some of our previous data sets (\cite{zho09b,ste10a,poe12}). 

To ensure adequate accuracy of the simulations, we compare the number of points we placed in the thermal boundary layer with the minimum number that should be placed inside the boundary layer according to \cite{shi10}. For each Pr number this criterion is satisfied for the highest resolution simulation and/or checked with resolution checks. Once a simulation is properly resolved there is no dependence on the grid resolution as the grid dependent errors diminish. This is a strict resolution criterium as it is sensitive to the resolution in the entire domain and not just in the boundary layers. Apart from the boundary layer the azimuthal resolution close to the sidewall has to be chosen properly in cylindrical domains. The only way to check this is to perform the same simulations on different resolutions and compare the results (\cite{ste10}). In order to check this we have performed the simulation for several Pr, especially the lowest and highest Pr, with different resolution and we find good agreement between the results obtained at different resolutions. We compared the average length scale in the flow with the largest grid scaling and the time convergence of the results. The results of this analysis are presented in the supplementary material.

\FloatBarrier
\section{Flow topology}
\label{topo}
\begin{figure}
\centering
\subfigure{\includegraphics[width=0.30\textwidth]{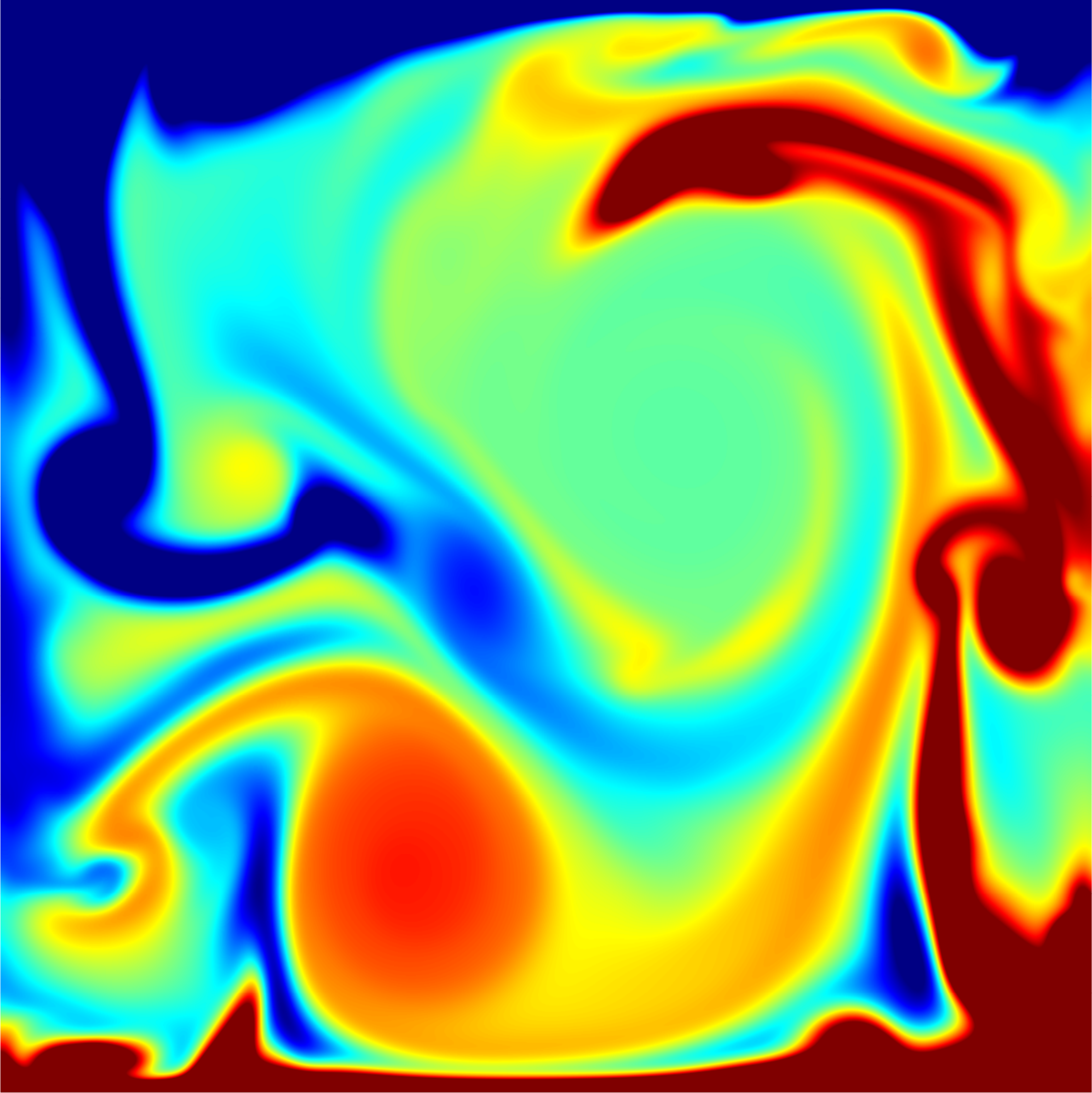}}
\subfigure{\includegraphics[width=0.30\textwidth]{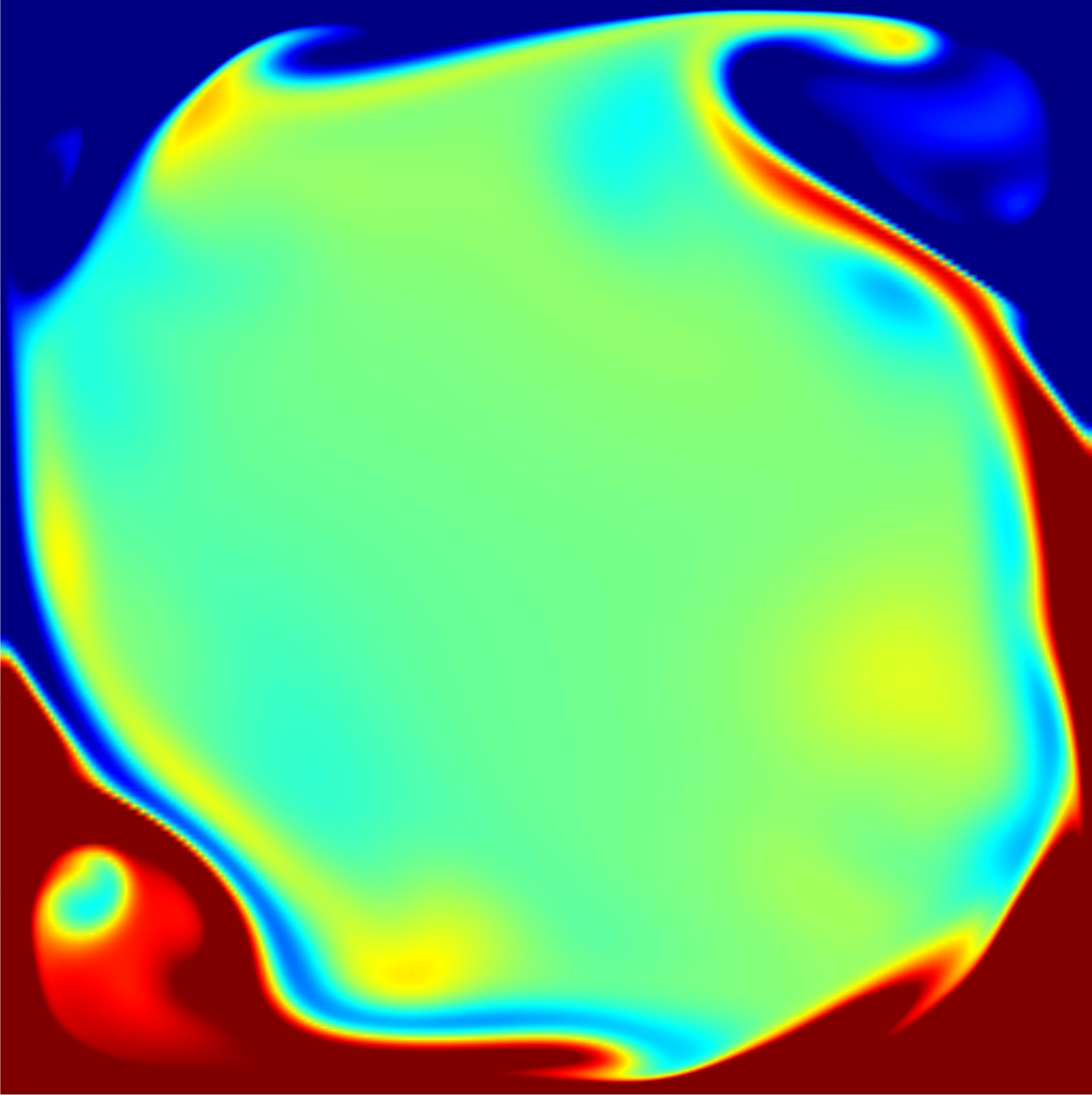}}
\subfigure{\includegraphics[width=0.30\textwidth]{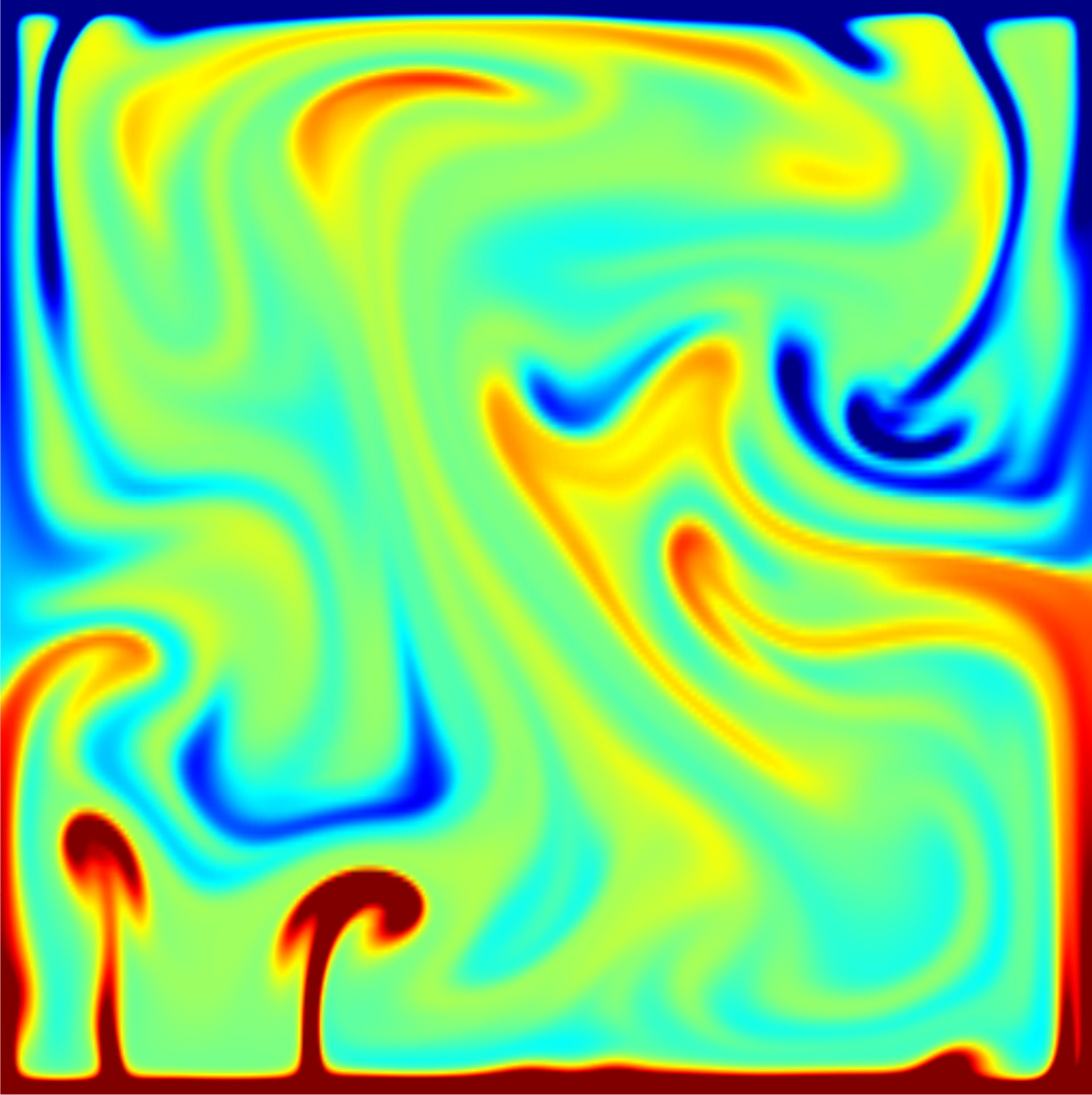}}
\subfigure{\includegraphics[width=0.30\textwidth]{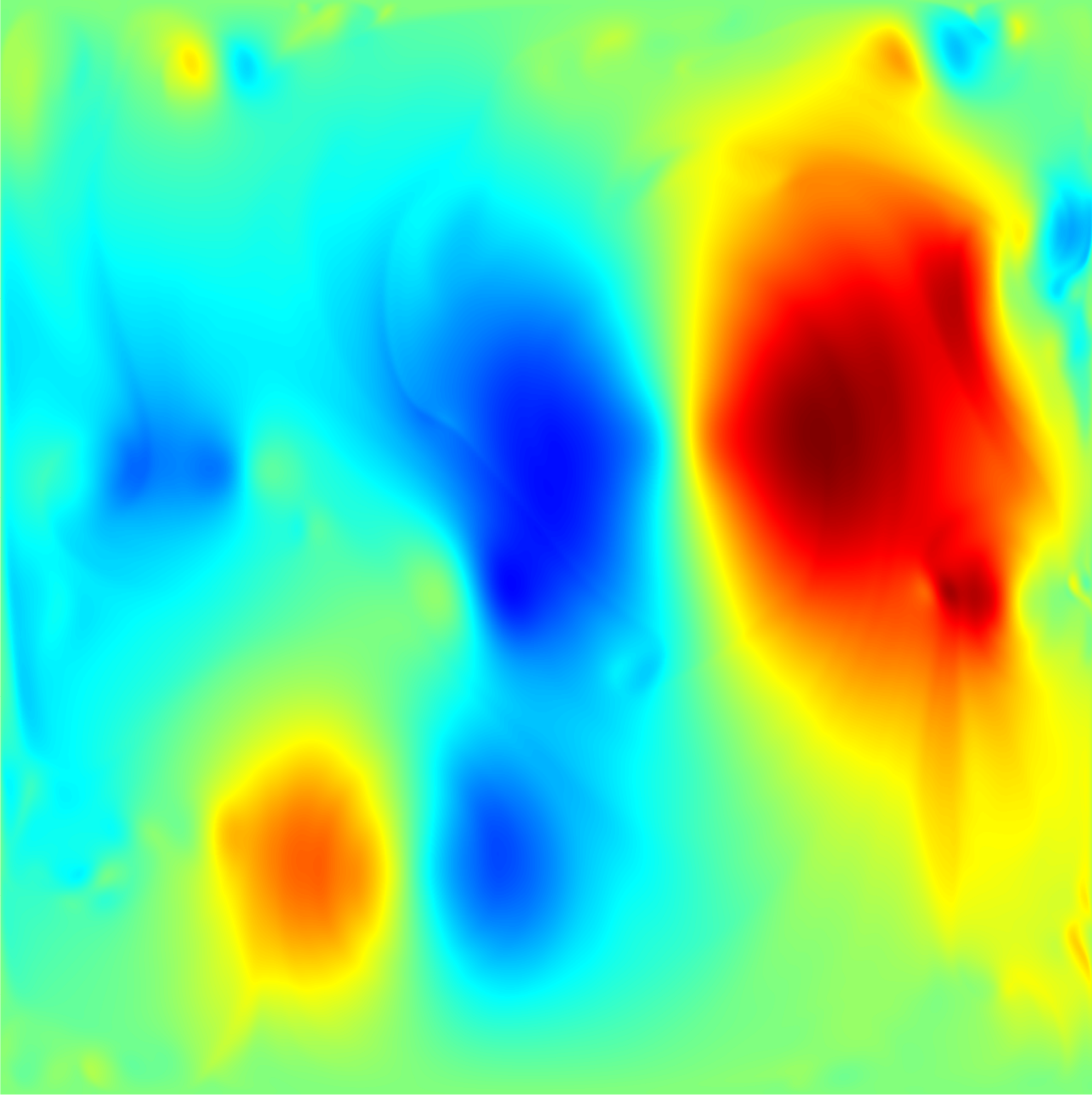}}
\subfigure{\includegraphics[width=0.30\textwidth]{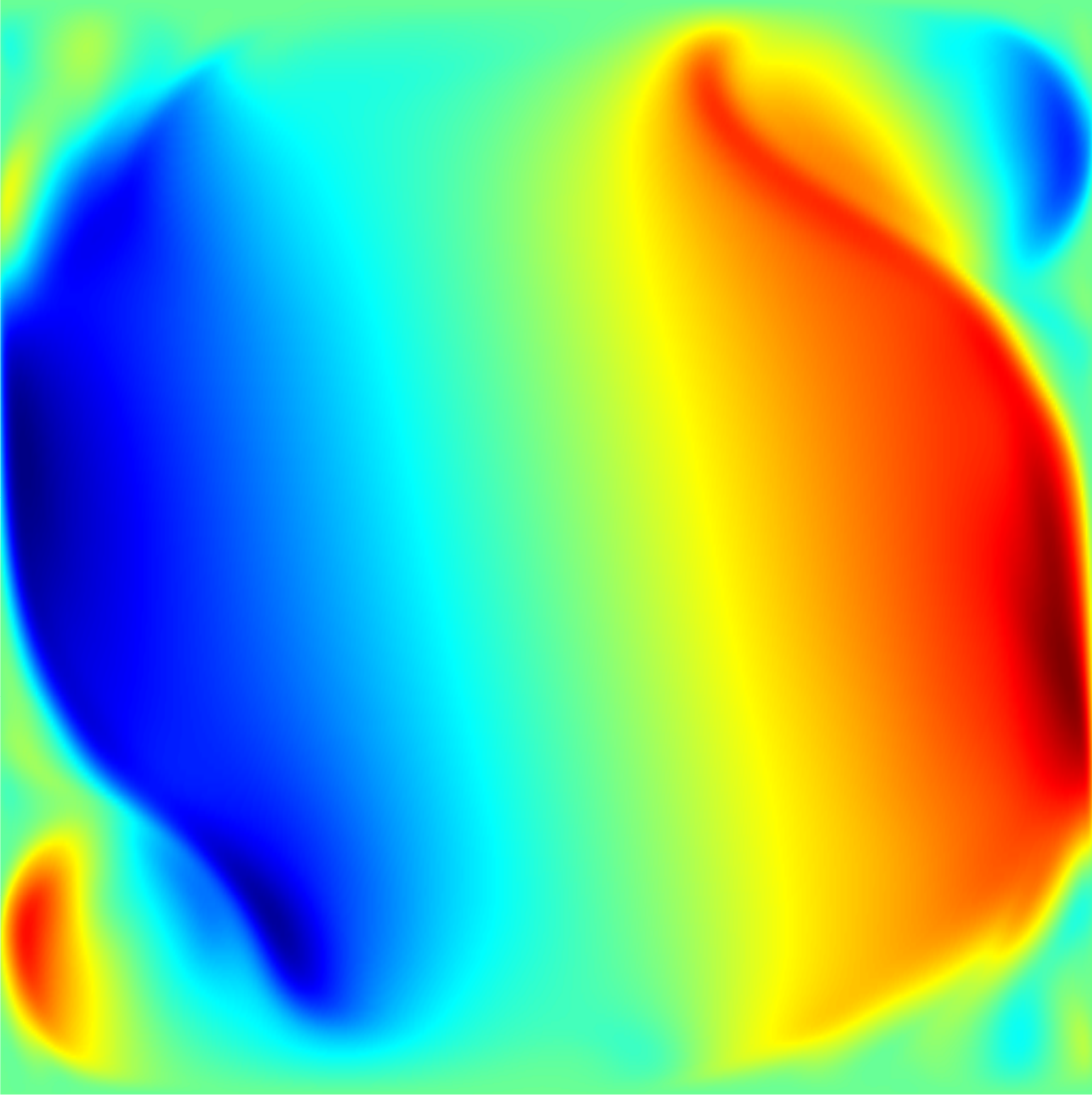}}
\subfigure{\includegraphics[width=0.30\textwidth]{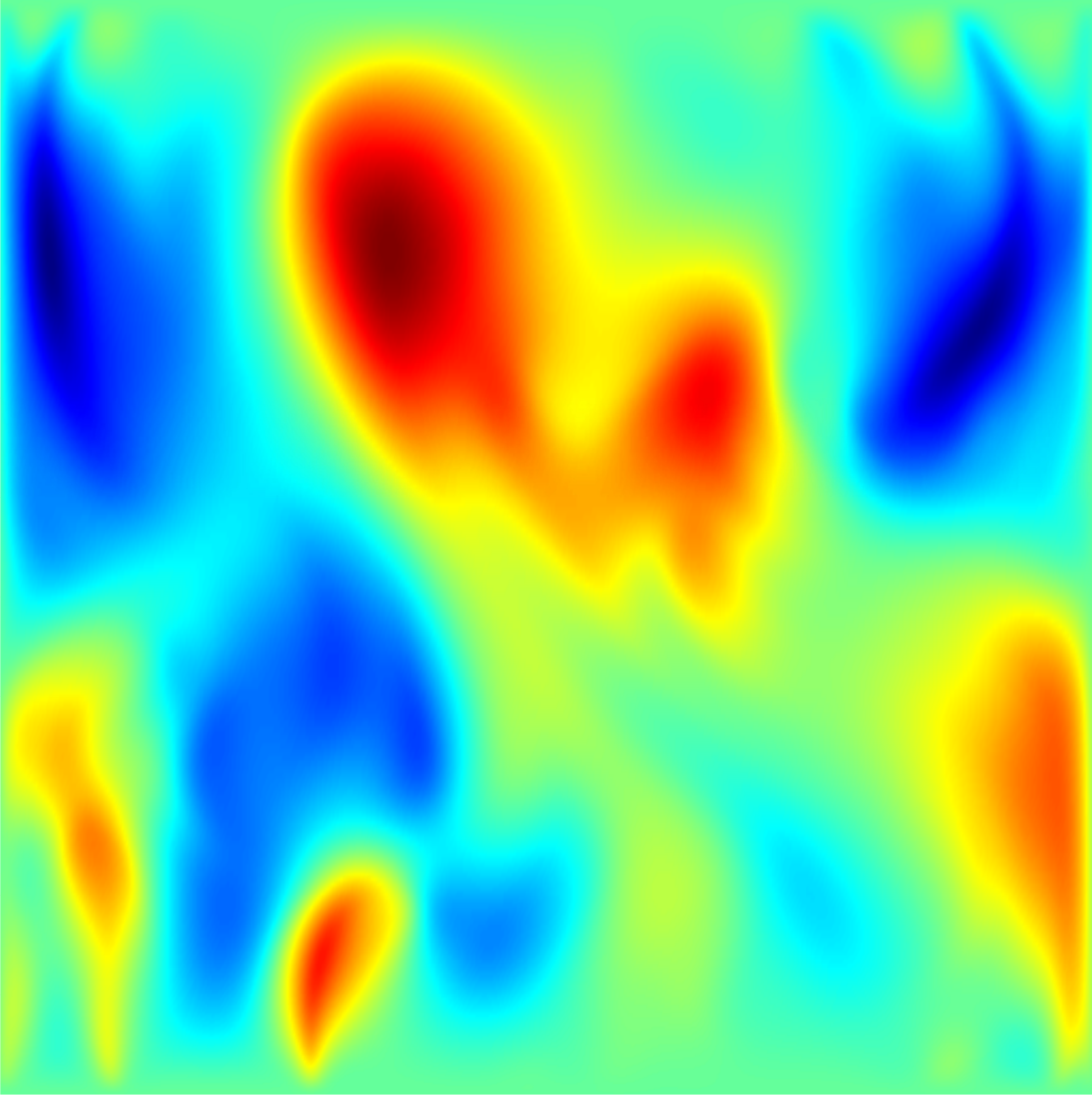}}
\caption{Temperature (top panels) and vertical velocity (bottom panels) of a 2D $\Gamma=1$ cell at $\tra=10^8$. The panels from left to right are for $\tpr=0.045$, $\tpr=0.7$, and $\tpr=55$. Red and blue indicate hot and cold fluid, respectively, for the temperature snapshots. For the velocity, red and blue indicate upward and downward moving fluid, respectively. The temperature colormap is the same for all temperature snapshots in figures \ref{fig:figure3b}, \ref{fig:figure3c} and \ref{fig:figure3} and ranges between $0.4 \leq \theta \leq 0.6$.}
\label{fig:figure3b}
\end{figure}

\begin{figure}
\centering
\subfigure{\includegraphics[width=0.30\textwidth]{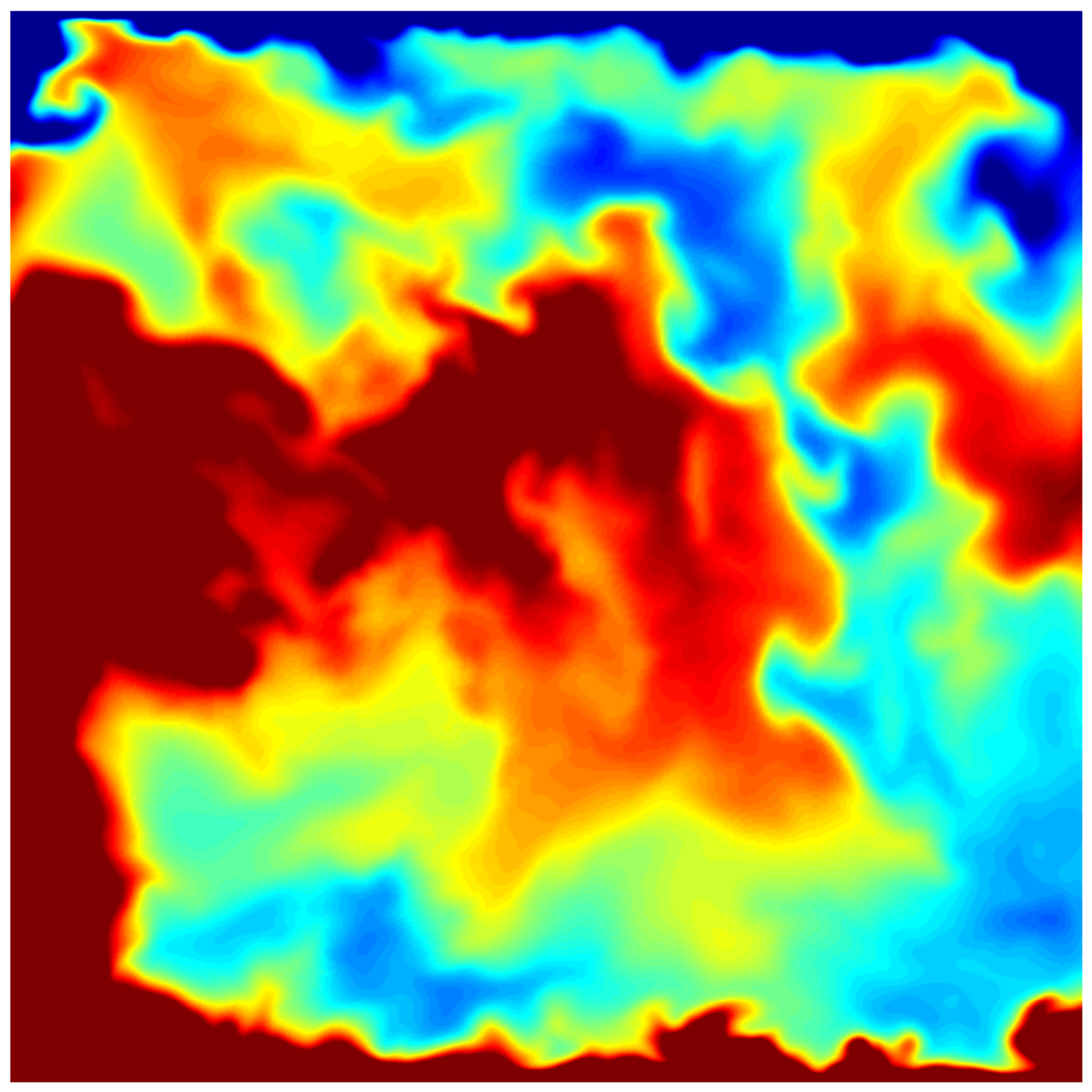}}
\subfigure{\includegraphics[width=0.30\textwidth]{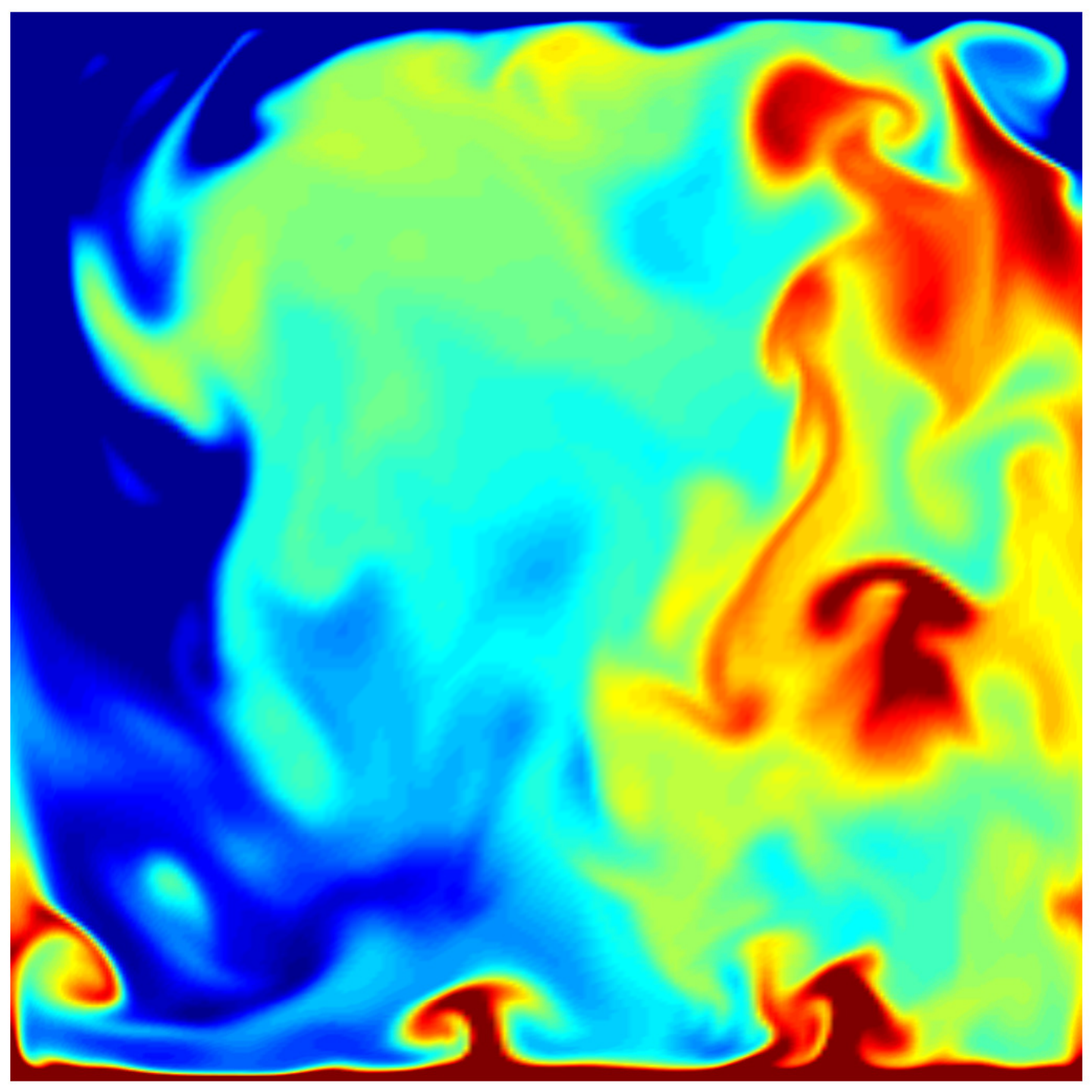}}
\subfigure{\includegraphics[width=0.30\textwidth]{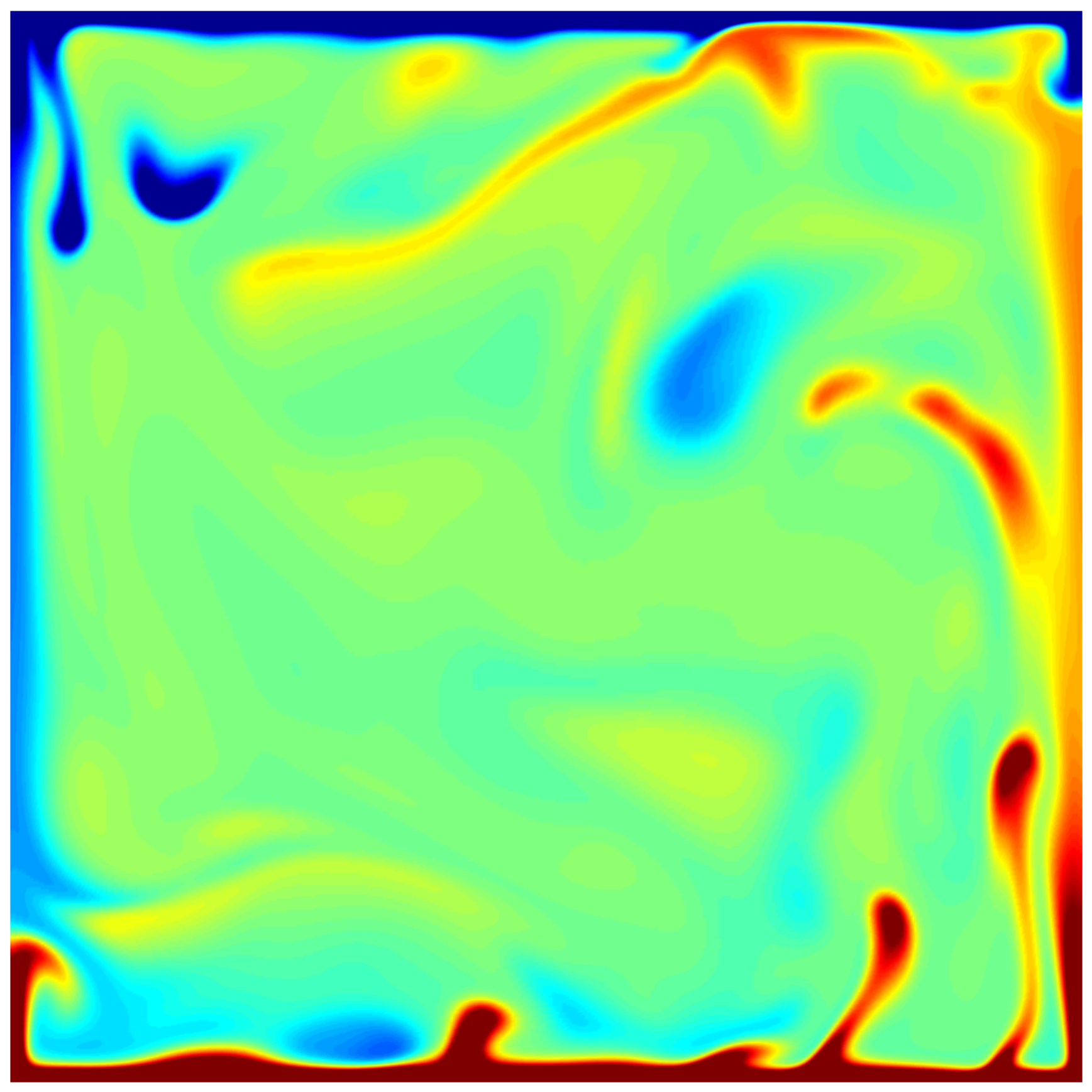}}
\subfigure{\includegraphics[width=0.30\textwidth]{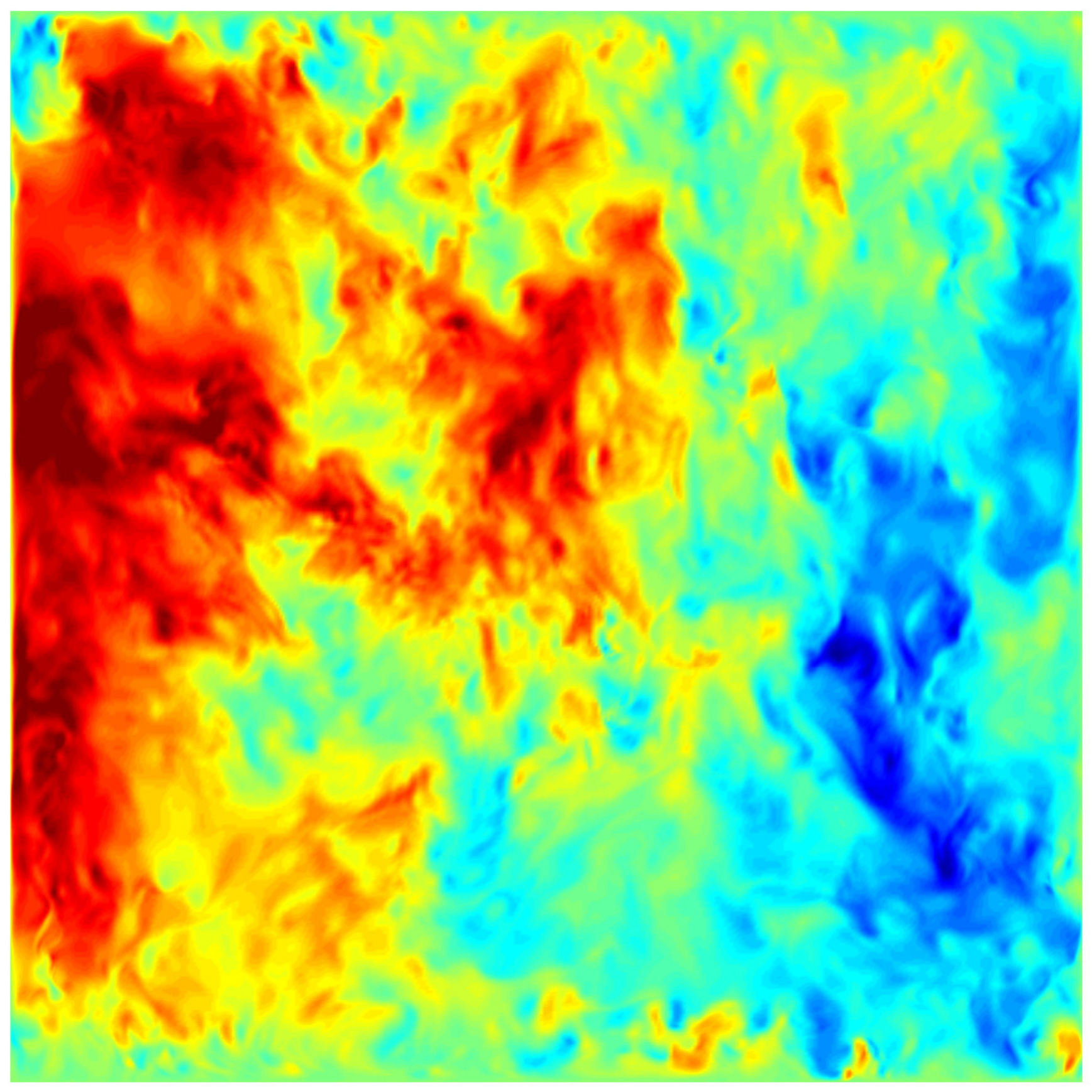}}
\subfigure{\includegraphics[width=0.30\textwidth]{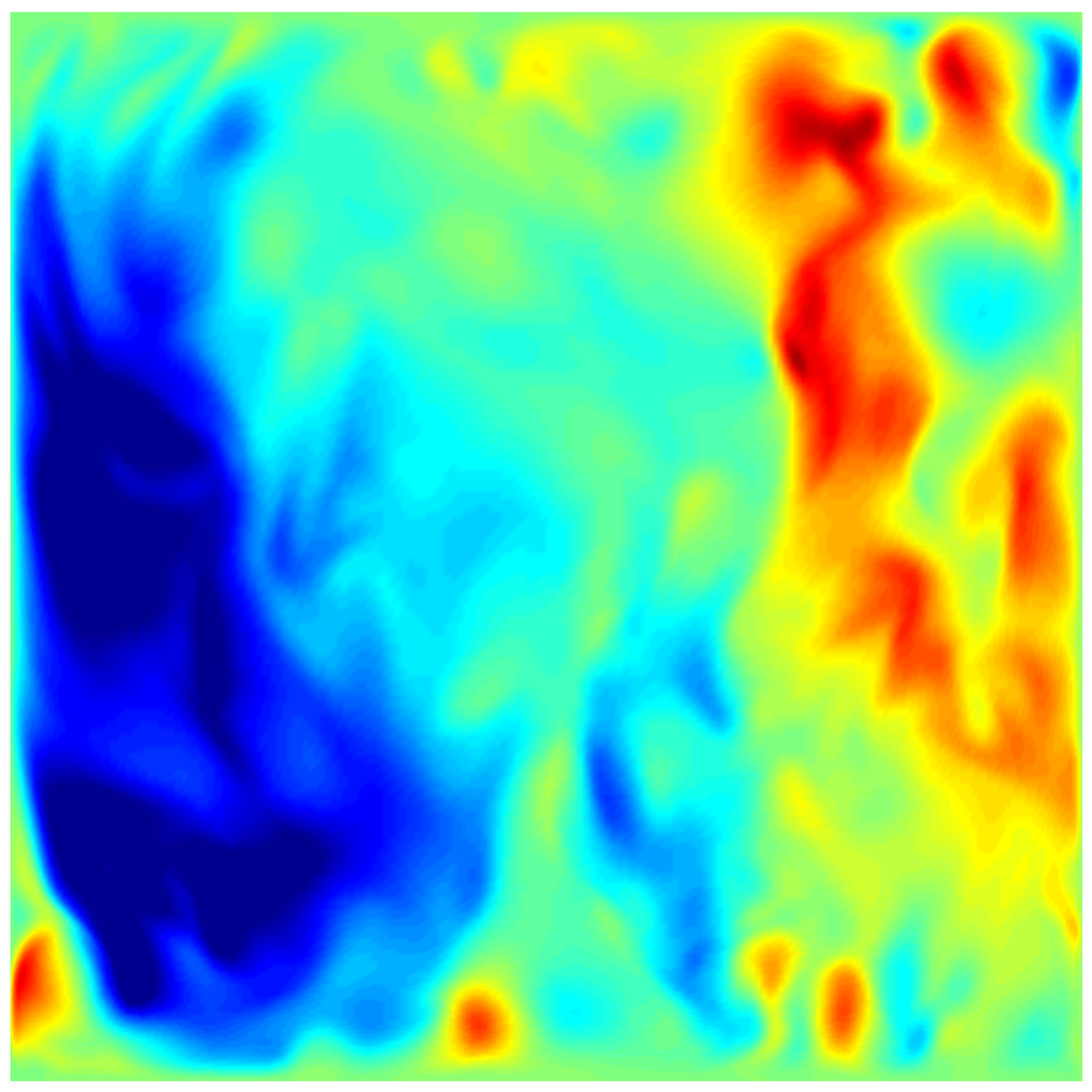}}
\subfigure{\includegraphics[width=0.30\textwidth]{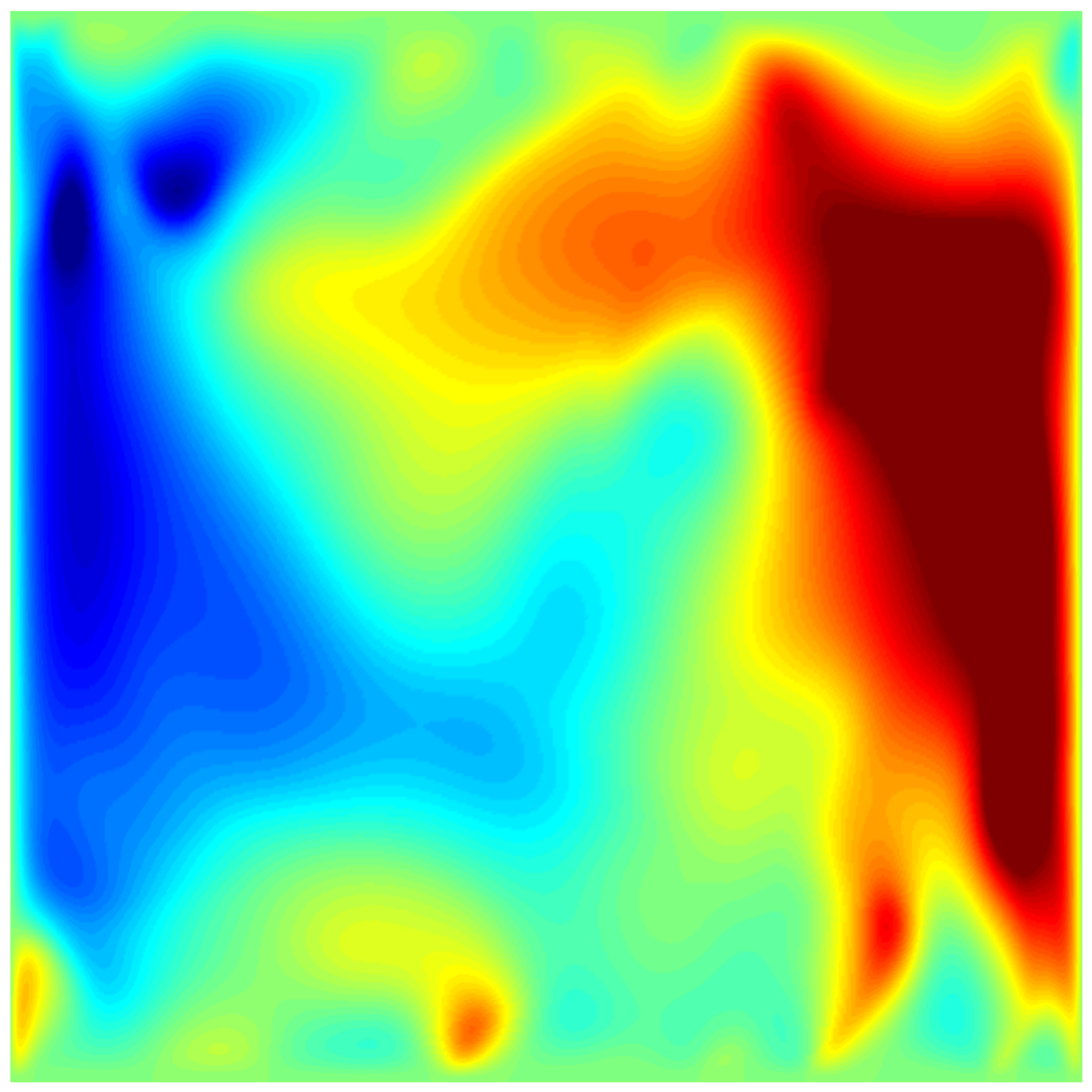}}
\caption{Temperature (top panels) and vertical velocity (bottom panels) at a vertical plane of a 3D cylindrical $\Gamma=1$ sample at $\tra=10^8$. The panels from left to right are for $\tpr=0.045$, $\tpr=0.7$, and $\tpr=55$. Same color coding as in figure \ref{fig:figure3b}. The azimuthal orientation of these vertical cross-sections can be seen in figure \ref{fig:figure3}}
\label{fig:figure3c}
\end{figure}

\begin{figure}
\centering
\subfigure{\includegraphics[width=0.30\textwidth]{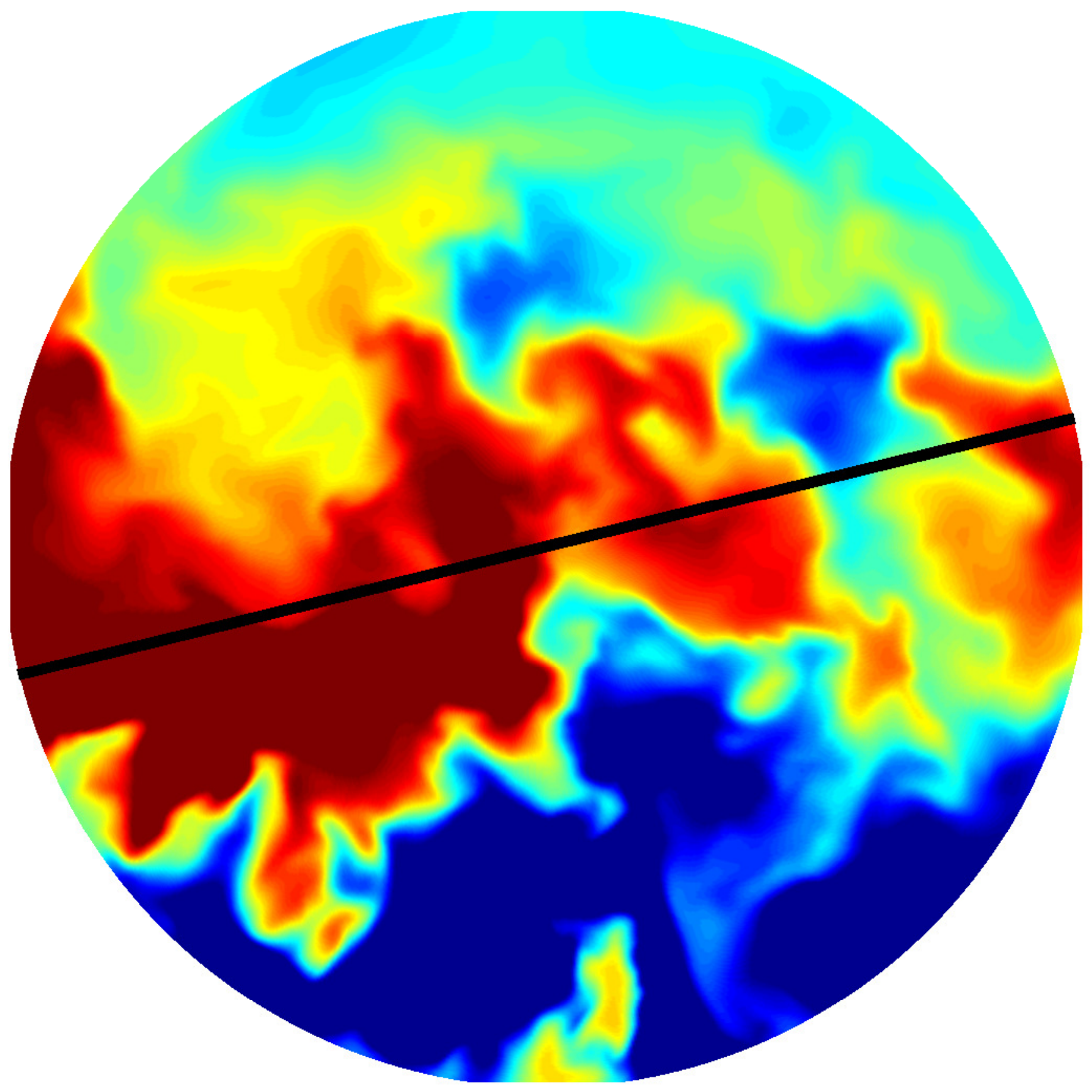}}
\subfigure{\includegraphics[width=0.30\textwidth]{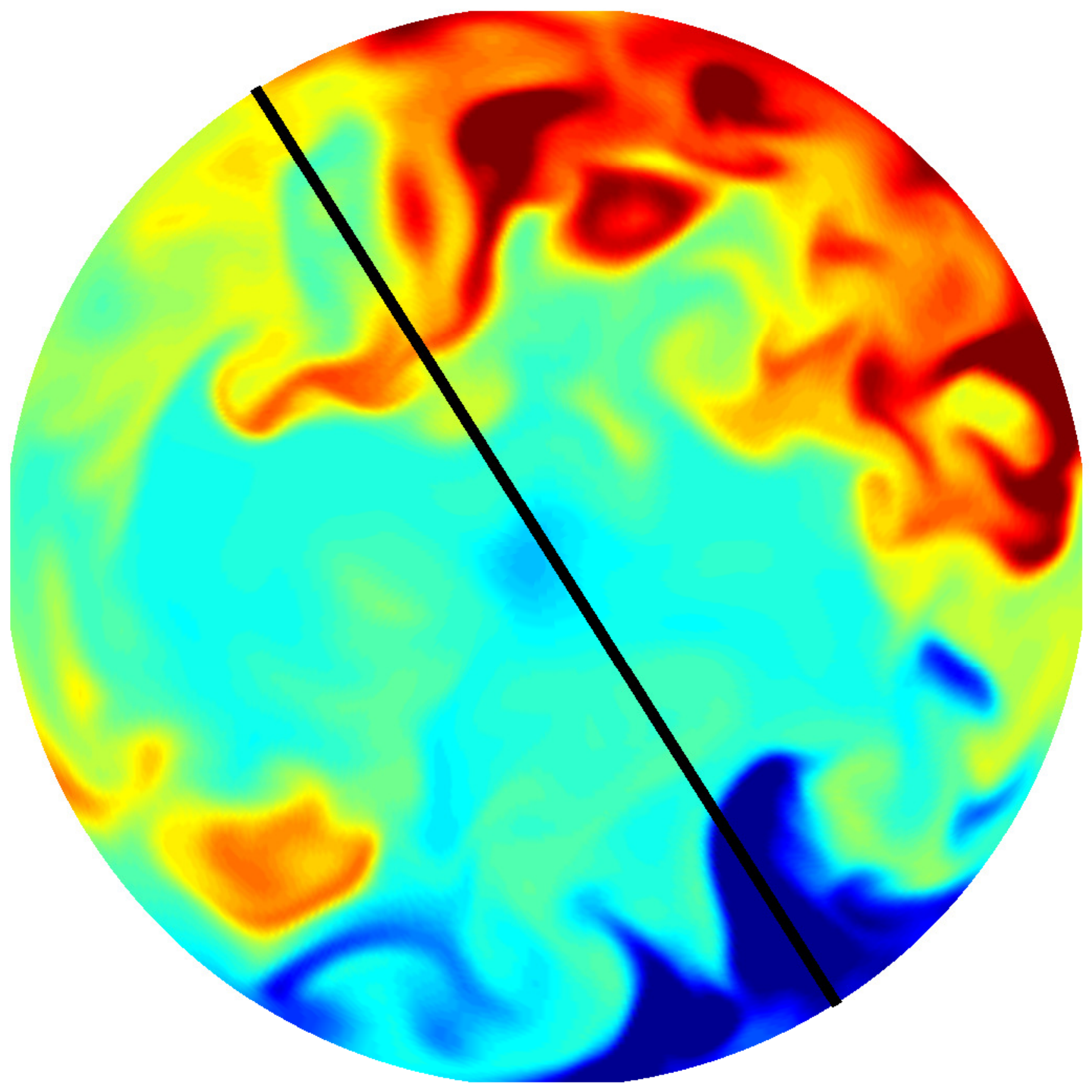}}
\subfigure{\includegraphics[width=0.30\textwidth]{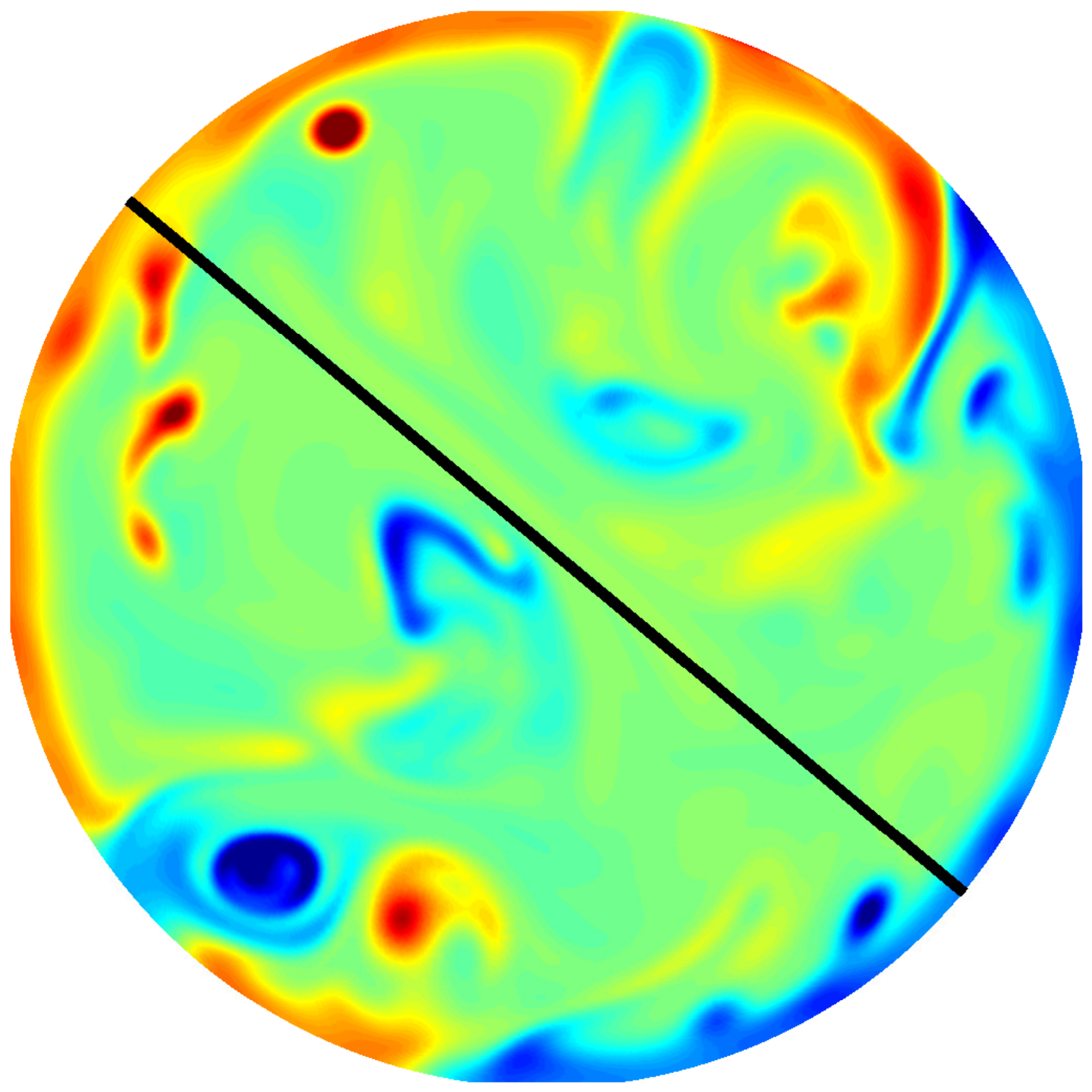}}
\subfigure{\includegraphics[width=0.30\textwidth]{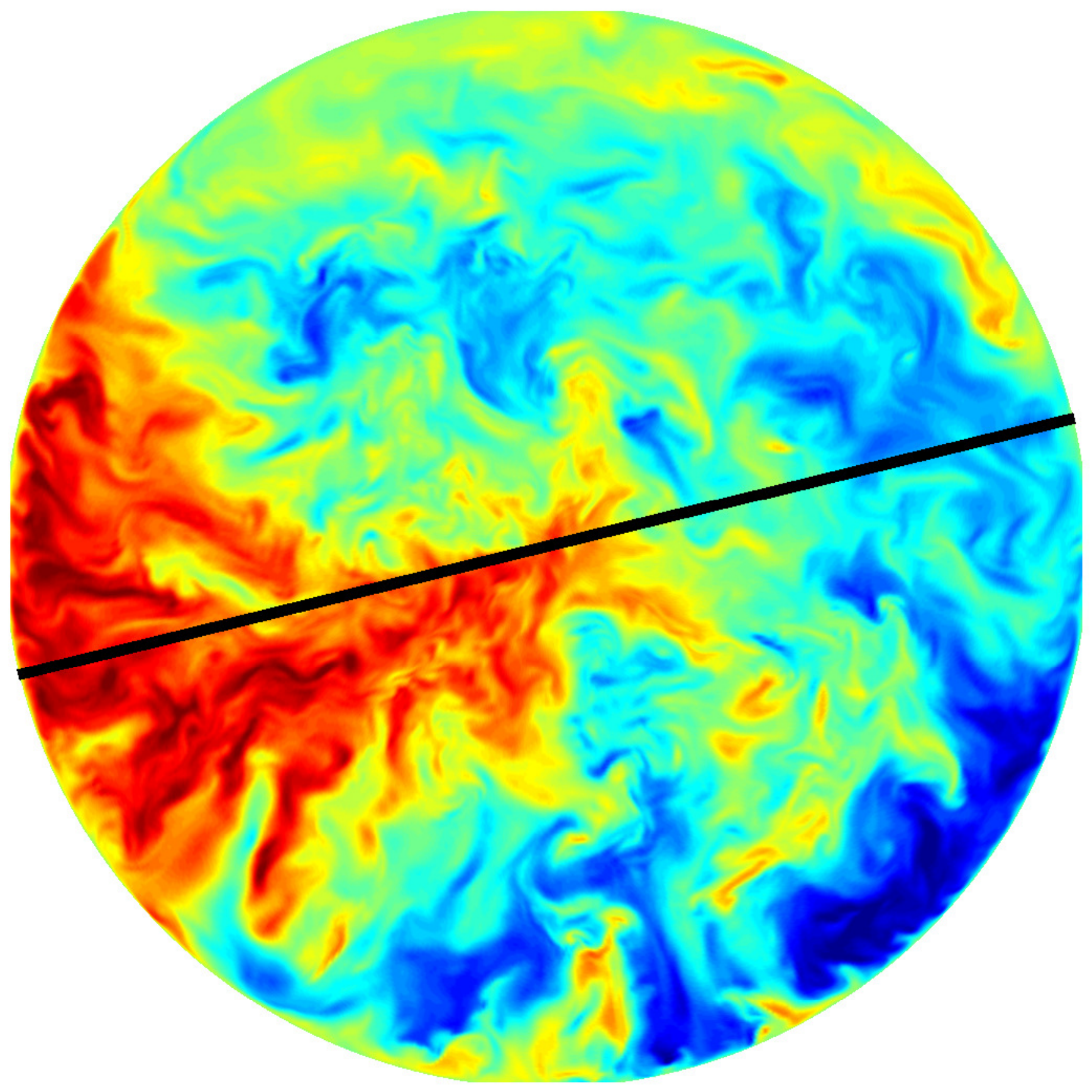}}
\subfigure{\includegraphics[width=0.30\textwidth]{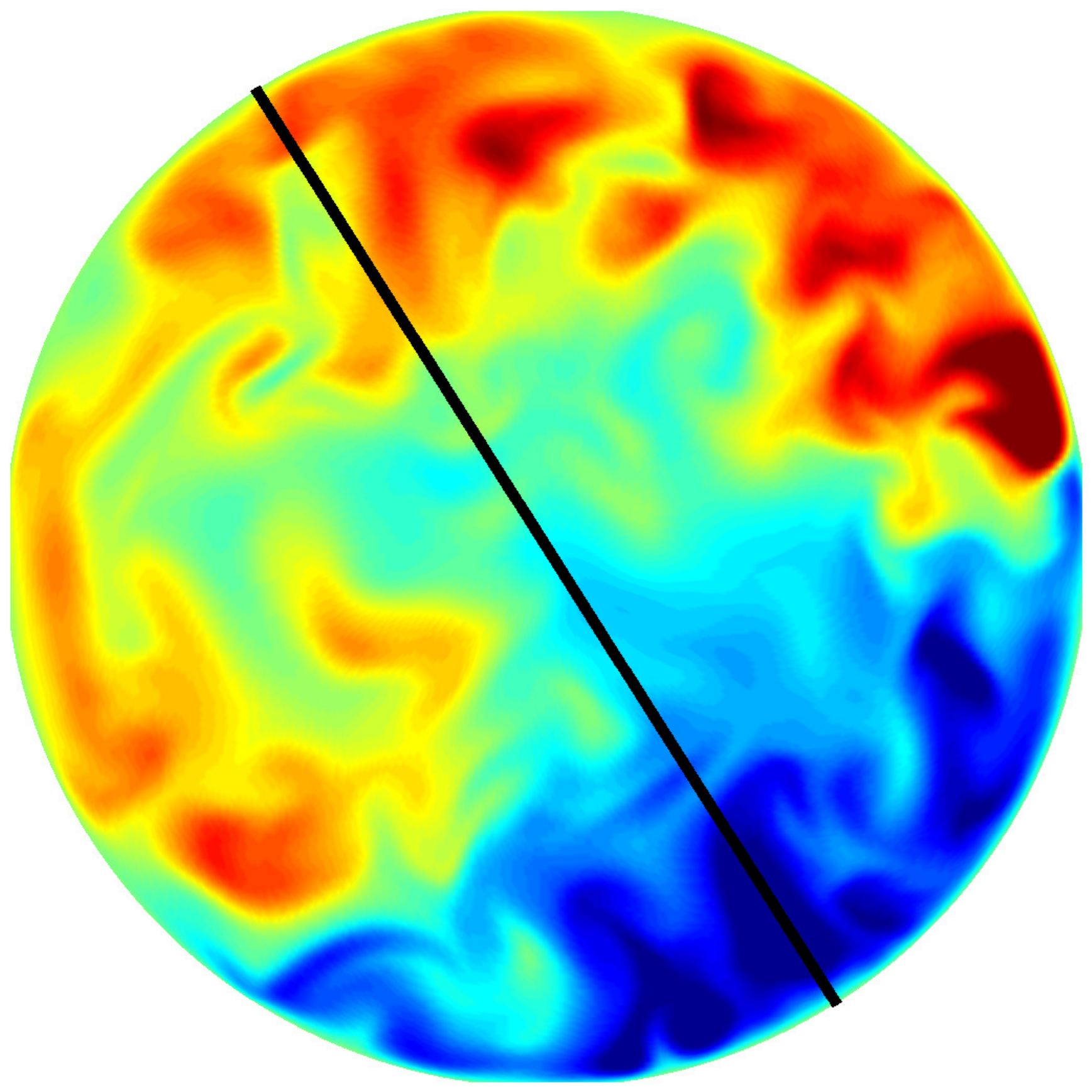}}
\subfigure{\includegraphics[width=0.30\textwidth]{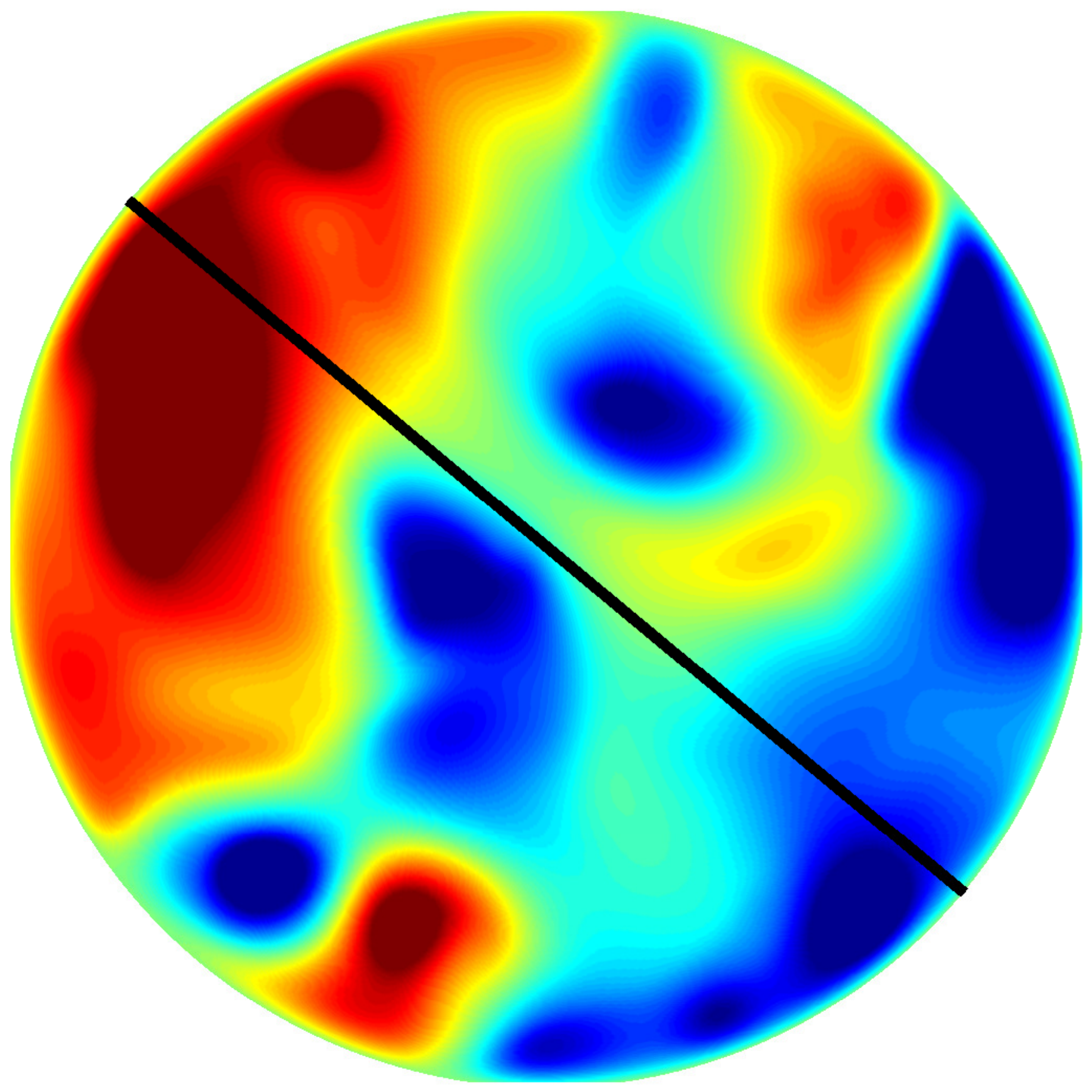}}
\caption{Temperature (top panels) and vertical velocity (bottom panels) at the horizontal midplane of a $\Gamma=1$ cell at $\tra=10^8$. The black lines indicate the azimuthal orientation of the corresponding vertical cross-sections found in figure \ref{fig:figure3c}. The panels from left to right are for $\tpr=0.045$, $\tpr=0.7$, and $\tpr=55$. Note that the velocity scales are visibly smaller than the temperature scales for the small Pr number case, while the temperature scales are smaller than the velocity scales for the high Pr number case.}
\label{fig:figure3}
\end{figure}

Figures \ref{fig:figure3b}, \ref{fig:figure3c}, and \ref{fig:figure3} show flow field snapshots of the complete 2D field, a vertical cross-section and a horizontal cross-section of the 3D cylinder, respectively. The top row of the panels depicts the temperature field and the bottom row depicts the vertical velocity field. The left, center, and right column are for $\tpr = 0.045$, $\tpr = 0.7$ and $\tpr = 55$, respectively. In the 3D panels it becomes apparent that the temperature structures become more localized at $\tpr = 55$, as is expected for high Pr flows. At high Pr there is hardly any LSC and the flow is plume-dominated (\cite{ver03}). This is reflected in 2D, for which the panels of both the temperature and velocity look very similar as in 3D, which is in agreement with \cite{sch04} who concluded that 2D and 3D is similar at high Pr due to the vanishing toroidal component of the velocity. Another interpretation of the similarity of the flow topology at high Pr can be made using the plume topology, as at high Pr the flow is plume dominated. In 2D it can be seen that for increasing Pr, the plumes change from roll-up type to sheet-like and finally to mushroom type. The roll-up type plumes are vortices that become buoyant by extracting thermal energy from the BL. These can be seen in figure \ref{fig:figure3b} for $\tpr = 0.045$. The sheet like plumes are elongated BLs stretching upwards and can be found for moderate Pr. For high Pr the flow is dominated by mushroom shaped plumes. One can imagine that 3D mushroom type plumes can be reduced to 2D through axisymmetry, while the other types do not possess symmetry that translates from 2D to 3D without violating the divergence-free condition imposed on the velocity field and/or the no-slip boundary conditions. For example, the 3D analogue of a roll-up plume would be a cylinder.

The visual differences emerge at $\tpr < 1$. At $\tpr = 0.7$ a pronounced LSC with corner rolls can be seen in 2D in figure \ref{fig:figure3c}. In 3D, the LSC is less pronounced and the corner rolls are much smaller. These differences might be due to the absence of a preferential azimuthal orientation of the LSC in 3D (\cite{fun04,bro05,xi06}). The azimuthal orientation of the vertical cross-sections displayed here is selected to obtain the most clear depiction of the LSC. In addition, it can be seen that in 3D, thermal plumes are emitted from the horizontal center of the boundary layer and move through the bulk, in contrast with 2D. This is due to the fact that in 3D, the LSC cannot fully enclose the flow and limit the movement of plumes.

A clear difference between 2D and 3D can be found at $\tpr = 0.045$. In particular, the vertical velocity snapshots reveal a drastically different structure. The 2D field has locally very small velocity structures similar to the 3D field. However, even though both 2D and 3D appear to have a LSC, the average velocity scale in 3D is much smaller than in 2D. \cite{bur03} concluded that for $\tpr = 0.1$ and $\tra = 10^8$ the LSC is driven by buoyancy forces more than by small scale turbulent fluctuations in both 2D and 3D. While the small scales do appear to have merged with the LSC in 2D, they are possibly not the dominant contribution to the driving of the LSC.

\section{Nusselt number}
In this section we will first compare the Rayleigh number scaling of the Nusselt number obtained in 2D and 3D simulations before we compare the Prandtl number dependence in detail.
\subsection{Rayleigh number dependence}
\begin{figure}
\centering
\subfigure{\includegraphics[width=0.49\textwidth]{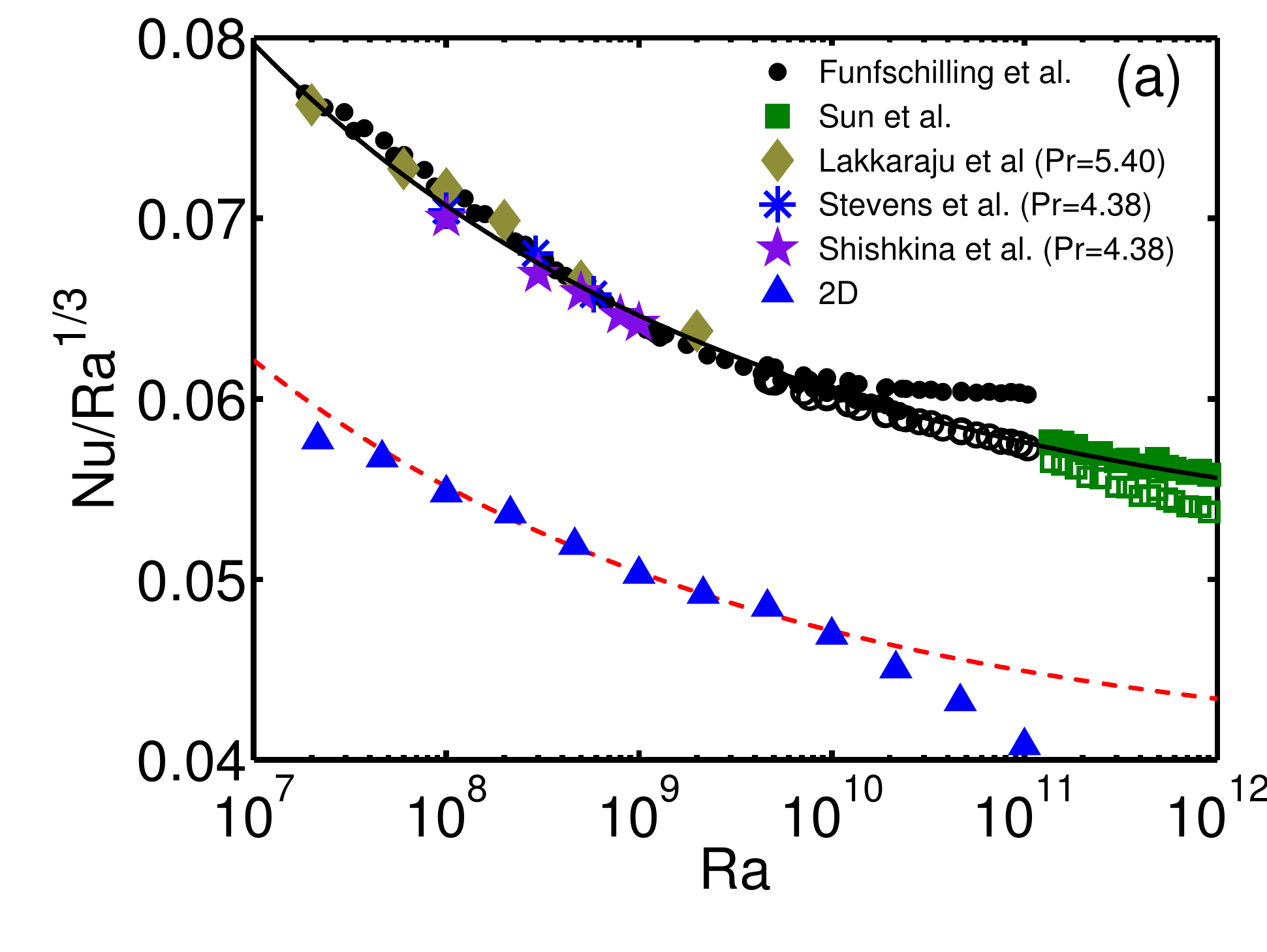}}
\subfigure{\includegraphics[width=0.49\textwidth]{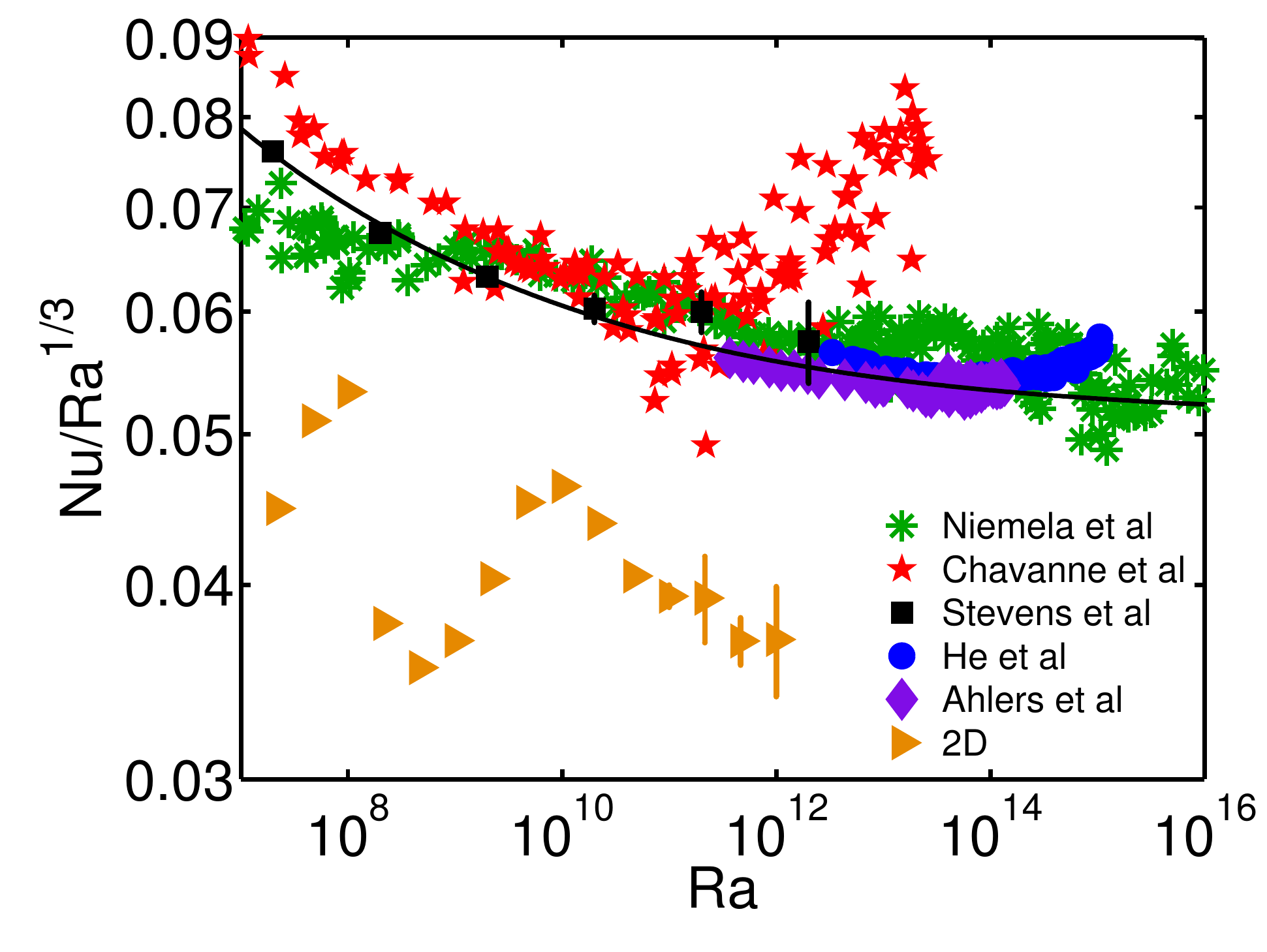}}
\caption{a) Nu versus Ra scaling for water ($\tpr=4.38$) in a $\Gamma=1$ sample. The simulation result for 2D RB are indicated by the blue upward pointing triangles. The 3D experimental results from \cite{fun05} and \cite{sun05d} are indicated by circles and squares. respectively. Here the open symbols indicate uncorrected data, while the filled symbols indicate corrected data. The numerical results from \cite{shi09}, \cite{ste11b}, and \cite{lak12} ($\tpr=5.4$) are indicated by the purple stars, blue asterisks and olive diamonds, respectively. The black solid line indicates the GL prediction and the red dashed line indicates the GL prediction multiplied by a constant value $A$ of $0.78$. The rest of the points are 3D experimental and numerical data. b) Nu vs Ra for $\tpr = 0.7$ in a $\Gamma = 0.5$ sample. The 2D data are indicated by the yellow rightward pointing triangles. The stars (\cite{cha01}), asterisks (\cite{nie00}), diamonds (\cite{ahl09c}) and circles (\cite{he12,ahl12b}) indicate experimental data and the squares results from numerical simulations (\cite{ste10,ste10d}). The refitted GL theory is indicated by the black line.}
\label{nura}
\end{figure}

In figure \ref{nura}a the compensated Nusselt number $\tnu/\tra^{1/3}$ as a function of Ra for $\tpr = 4.38$ and $\Gamma = 1$ is displayed. The data is taken from 3D experimental results of \cite{fun05} and \cite{sun05d}, 3D numerics of \cite{shi09}, \cite{ste11b} and \cite{lak12} and 2D numerics of this research. Both the uncorrected and corrected experimental data is depicted. The corrected data is compensated for finite plate conductivity, see \cite{ahl09}. In these and upcoming figures, the 2D data is represented by triangles with varying orientations and the 3D data by other symbols. For reference, the refitted GL prediction for 3D is included. As the 3D data and GL theory display near equal results for the evaluated Ra range, the latter can be used as a guide in comparing the 2D Nu(Ra) data with 3D by rescaling it with a constant factor of 0.78. For 3D and 2D the scaling of Nu(Ra) agree very well for $10^7 \leq \tra \leq 10^{10}$. At higher Ra the 2D points are smaller than the rescaled 3D GL prediction. This indicates that for $\tra>10^{10}$ the 2D scaling differs from 3D as the 3D data does follow this scaling. An analysis of the roll states of $\tra = 4.64\cdot 10^9$ and $\tra = 10^{10}$, reveal that there is a substantial change in flow state between these Ra, which might be connected to the discrepancy in scaling. At $\tra = 4.64\cdot 10^9$, the flow is in a singe roll state similar to the state depicted for $\tpr = 0.7$ and $\tra = 10^8$ in figure \ref{fig:figure3b} while at $\tra = 10^{10}$ the roll state has become uncondensed. Here, the term uncondensed signifies that there is no energy pile up at a scale close to system size and thus there is no LSC. The largest scale in the flow consists of two mobile and orbiting smaller rolls. That Nu is (counter-intuitively) lower for this broken LSC has been observed previously by \cite{poe12} for $\tpr = 0.7$ and $\tra = 10^9$ and we believe that this is due to the increased path length of the thermal plumes before they can deliver the heat to the opposite plate. In case of a LSC the plumes move directly from their original BL to the opposite BL, while otherwise the plumes move less directly to the opposite BL, interacting with the multiple rolls composing the bulk. Now, not only the absolute Nu but also the scaling between Nu and Ra has appeared to be lower for this roll state. Extrapolating towards higher Ra, one expects that the scaling will change subsequently when these orbiting rolls are replaced by a more complex roll state with even smaller scales. Eventually, the fluctuations will become too large and the scales too small for a coherent roll state to exist that can affect integral quantities. In \cite{poe12} we showed that the scaling of Nusselt can change locally in 2D RB convection and can recover to the expected 3D scaling for higher Ra, see also the $\Gamma=1/2$ results. 

In 3D no such transition in an integral quantity exists as the LSC does not fully enclose the system. This gives the thermal plumes more freedom to move from one boundary layer to another. Therefore, the difference in Nu between a system with a single roll state and with a broken single roll state is expected to be small and more gradual than in 2D, where the system can jump between these states, affecting Nu (\cite{poe12}) and its scaling. 

The difference between 2D and 3D is expected to be larger for lower Pr due to the larger toroidal component of the velocity. In addition it is known from \cite{poe11} that the integral quantities and flow state in 2D have a stronger dependence on the aspect-ratio $\Gamma$ than in 3D (\cite{bai10}). This is emphasized by the increased effect of the flow state on Nu for low Pr due to the thermal boundary layer being exposed to the bulk flow (\cite{poe11}). For low Pr, the bulk flow directly extracts heat from the thermal BL and therefore the flow state of the bulk has a large effect on Nu. We therefore include a comparison for $\tpr = 0.7$ and $\Gamma = 0.5$, where we expect a substantial difference. The result is displayed in figure \ref{nura}b. Although the average scaling exponent appears to be similar, the 2D data reveals much more structure than 3D. This is caused by multistability of different flow states and the large difference in Nu between these flow states. By increasing Ra, the system is successively in a triple roll state, an unstable triple roll state and back to a triple roll state, passing through an unstable region until the roll state breaks up. These roll states strongly affect the resulting Nu. This effect is expected to decrease as the LSC looses its strength (\cite{poe12}) and the length scales become smaller at higher Ra. 

\subsection{Prandtl number dependence}

\begin{figure}
\centering
\subfigure{\includegraphics[width=0.49\textwidth]{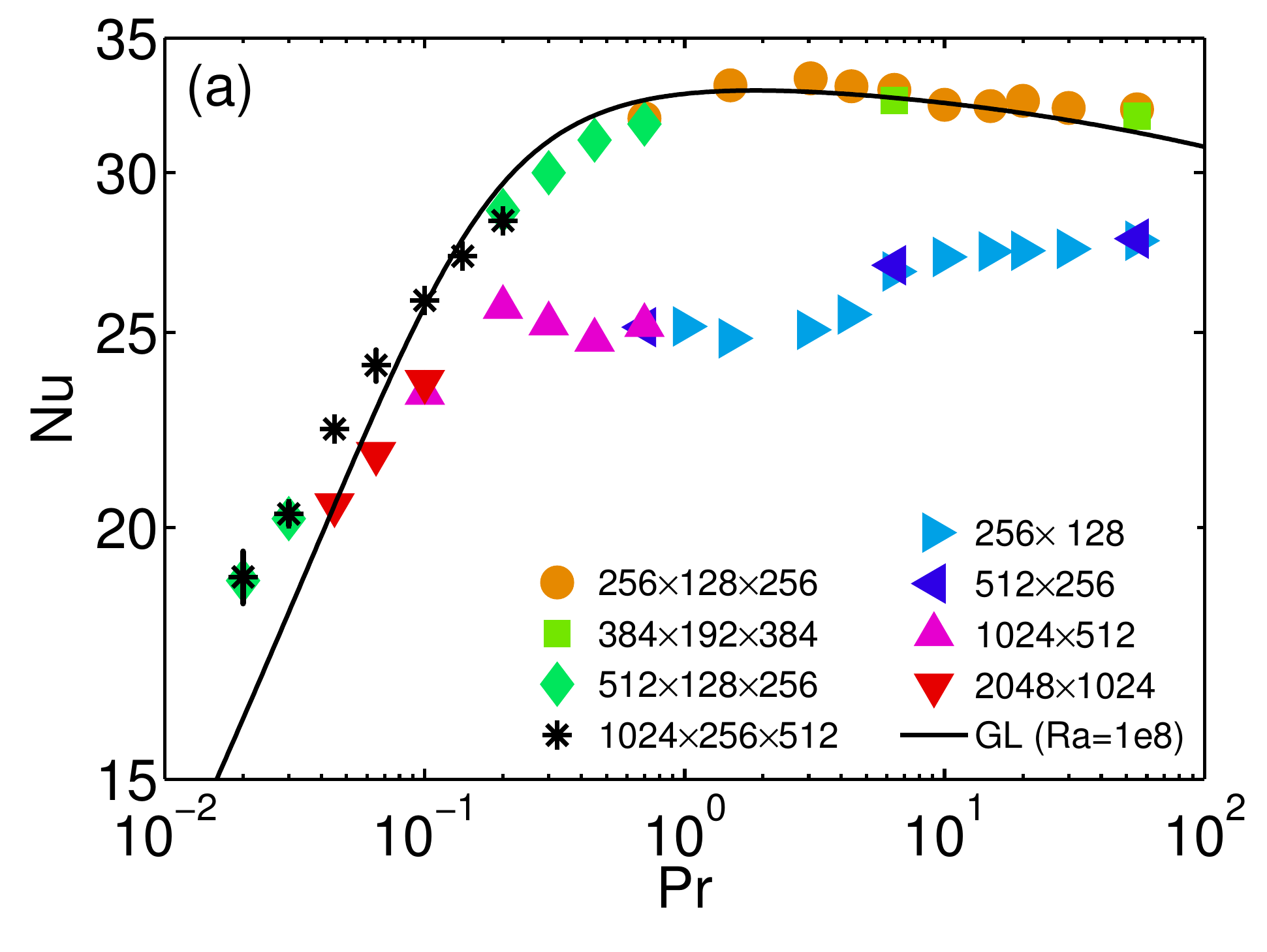}}
\subfigure{\includegraphics[width=0.49\textwidth]{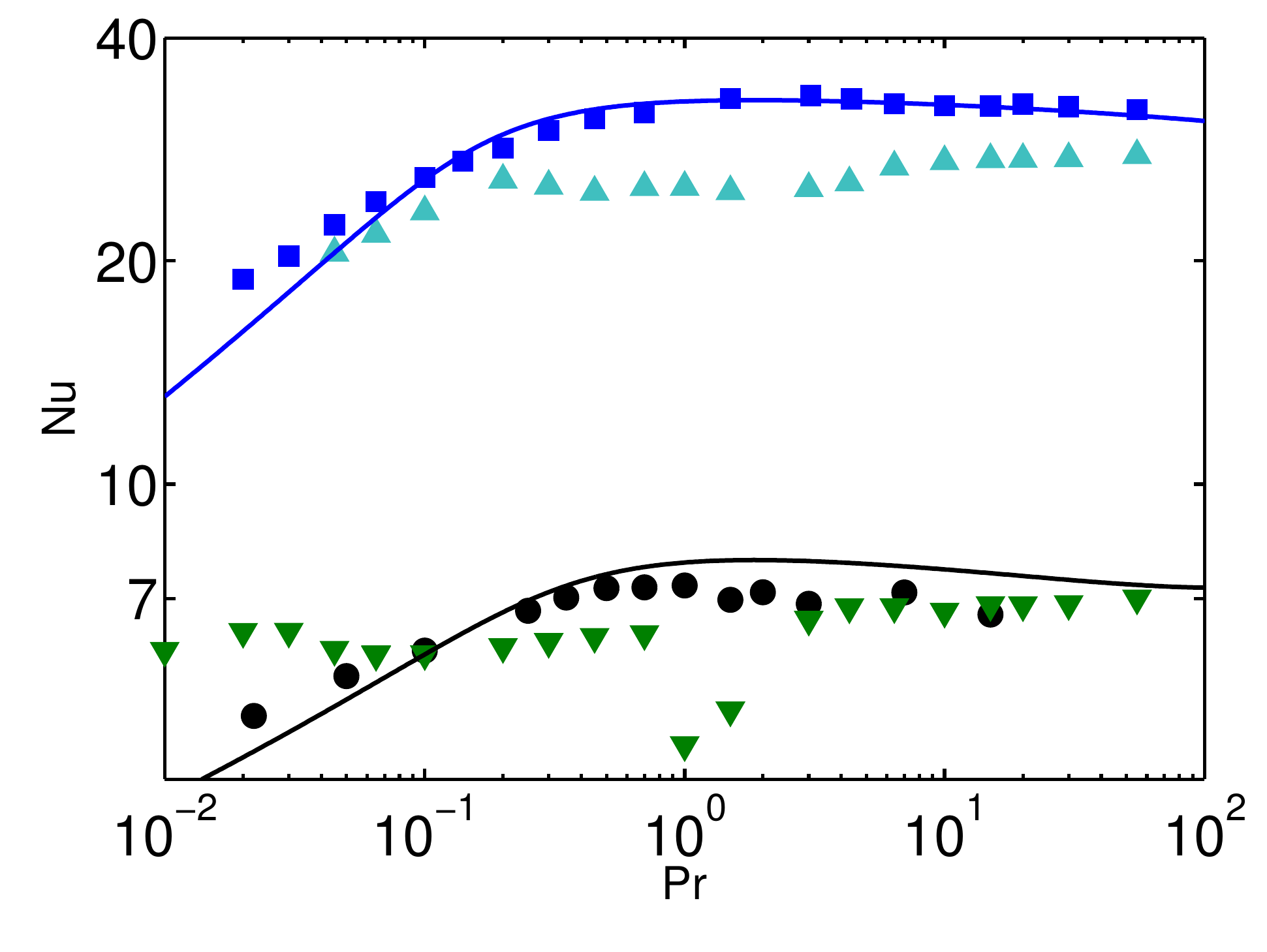}}
\caption{Nu as function of Pr. a) The Nusselt number obtained in numerical simulation performed on different grids for $\tra = 10^8$. The legend indicates the number of gridpoints in the vertical and horizontal direction for 2D and in the azimuthal, radial, and axial direction for 3D. b) Comparison of the Pr number dependence in 2D and 3D of Nu. In this panel the green and cyan upward pointing triangles indicate the results from 2D RB simulations at $\tra = 10^8$ and $\tra = 10^6$, respectively. For 3D this is indicated by the blue and black circles, respectively. The solid lines in both panels gives the prediction of the GL theory. 
}
\label{fig:figure2}
\end{figure}
The main conclusion from \cite{sch04} was that the agreement of global quantities between 2D and 3D depends on Pr. More specific, they conclude that for lower Pr the 2D output increasingly deviates from 3D. We repeat their measurements in 2D for $\tra=10^6$ and supplement it with a series for $\tra=10^8$, albeit with a no-slip boundary condition on the sidewalls in contrast with their stress-free sidewall boundary condition. In figure \ref{fig:figure2}a it can be seen that numerical simulations for low Pr become increasingly demanding in terms of resolution. Here the results for the $\tra=10^8$ runs are depicted with the symbol indicating the used numerical resolution. 
Figure \ref{fig:figure2}b shows Nu(Pr) for Ra for $10^6$ and $10^8$. The solid lines are the refitted theoretical GL predictions for the different Ra corresponding to the experimental and numerical data. First, we observe that the numerical results for 2D (green triangles) and 3D (black dots) at $\tra=10^6$ display no qualitative similarity, except for the Pr independence of Nu at higher Pr. Furthermore, in 2D multiple states are observed around $\tpr=1$, where the outlying points are caused by the double roll state of the system as opposed to the single roll state corresponding to the other data points. It is likely that the single roll state is stable as well, which would display a Nu similar to the surrounding Pr data (\cite{poe11}), however this is not checked. The discrepancy that \cite{sch04} did not observe these multiple states might be due to multistability or caused by their free-slip sidewall boundary condition. 
For $\tra = 10^8$ the 2D and 3D seem to converge for high Pr. The largest difference is seen at intermediate Pr, which is reflected in the flow topology, see section \ref{topo}. Here the strong LSC results in a substantial difference in Nu between 2D and 3D. At low Pr, unlike for $\tra = 10^6$, the 2D and 3D Nu are matching. However, the amount of data points is too low to make a strong conclusion on this surprising low Pr behavior.

\section{Reynolds number}

\begin{figure}
\centering
\subfigure{\includegraphics[width=0.49\textwidth]{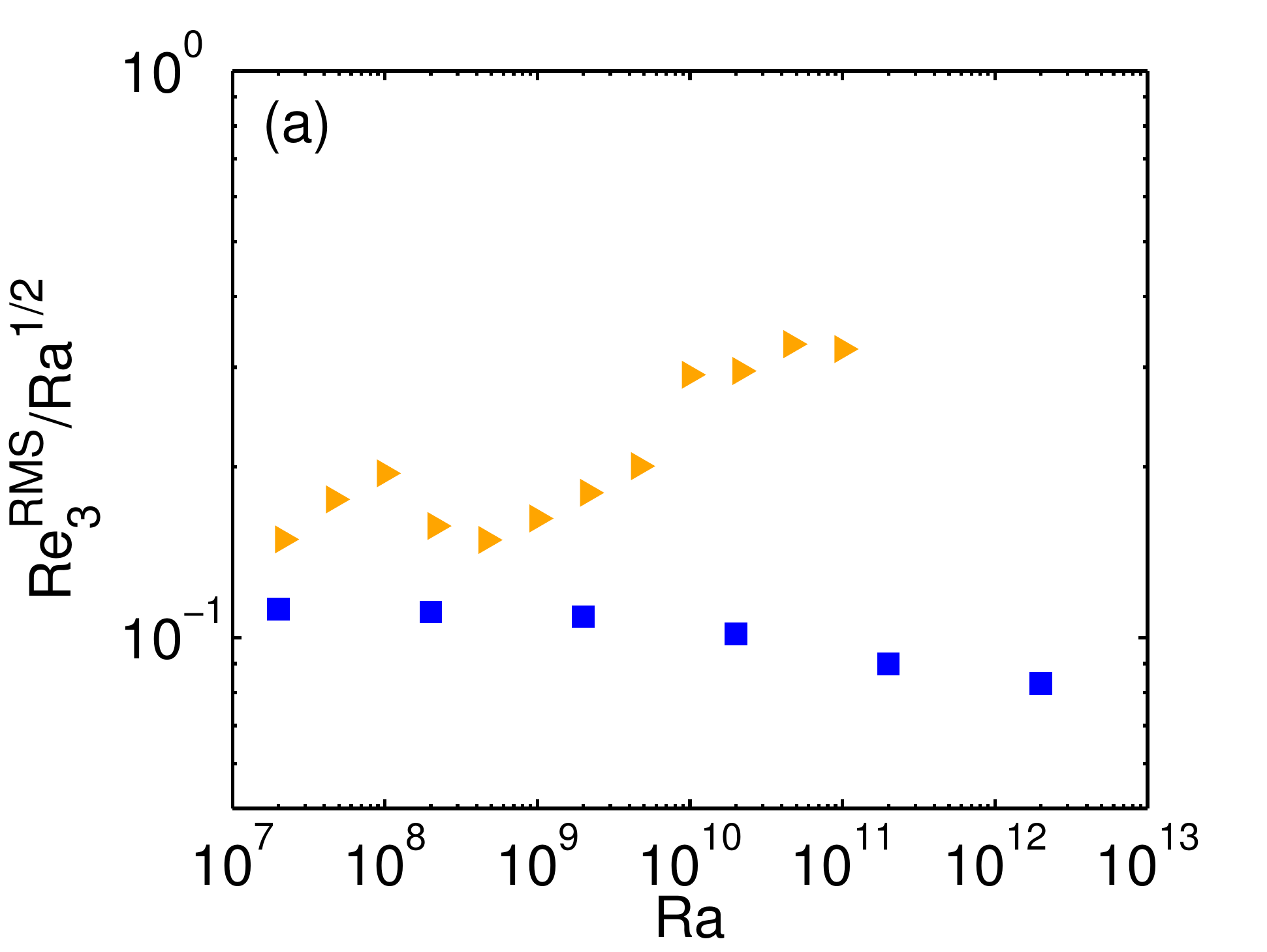}}
\subfigure{\includegraphics[width=0.49\textwidth]{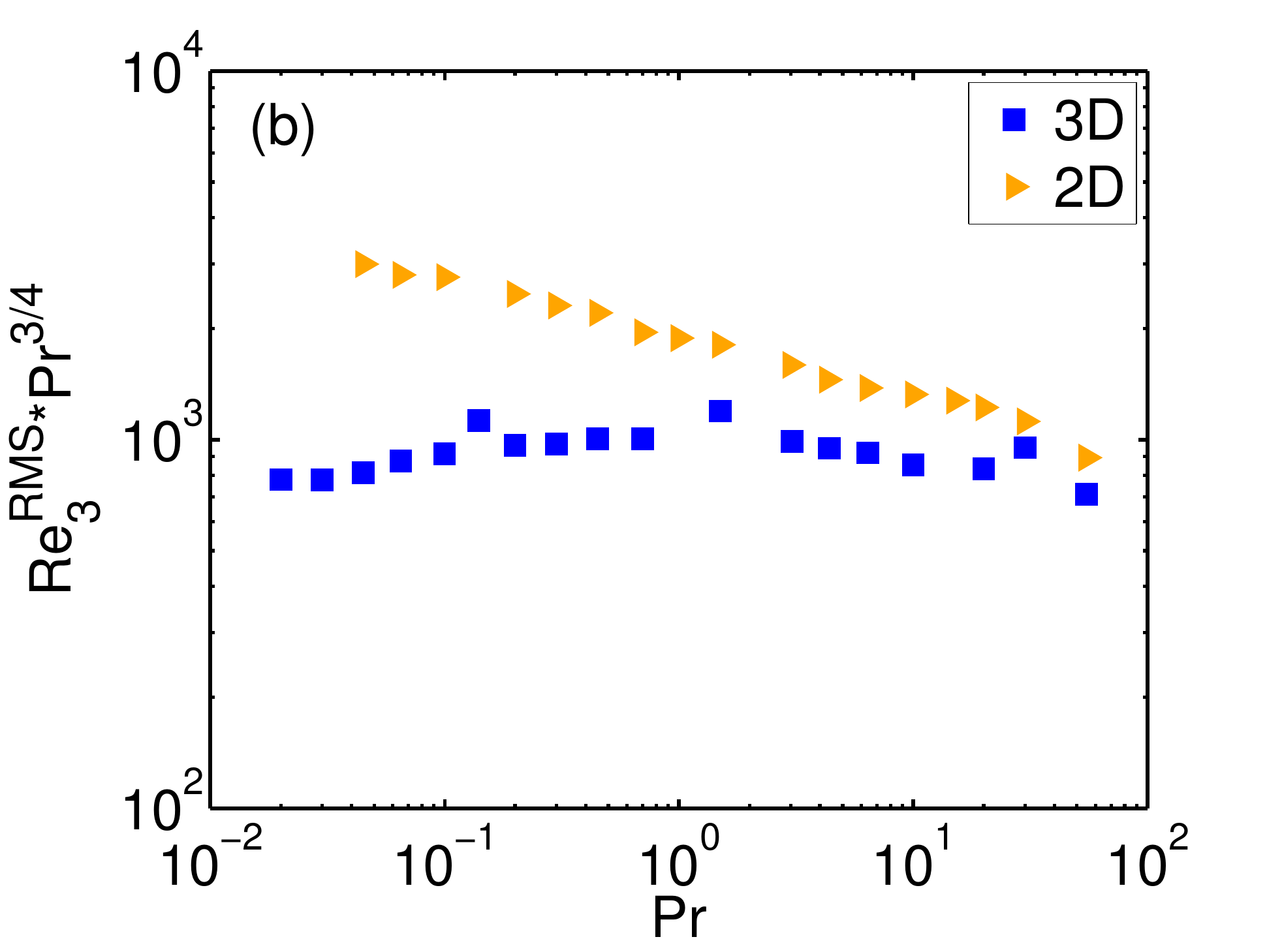}}
\caption{The compensated Reynolds number based on the root-mean-square vertical velocity $\tre_3^{RMS}$ for 2D (rightward pointing triangles) and 3D (squares). a) $\tre_3^{RMS}$ as a function of Ra for $\tpr = 0.7$ and $\Gamma = 0.5$. b) $\tre_3^{RMS}$ as a function of Pr for $\tra = 10^8$ and $\Gamma = 1$. The legend in b) applies to both plots. Corresponding Nu comparison can be found in figures \ref{nura}b and \ref{fig:figure2}b for a) and b), respectively.}
\label{fig:figureRe}
\end{figure}

Figure \ref{fig:figureRe} shows the comparison for the compensated Reynolds number based on the root-mean-square vertical velocity $\tre_3^{RMS} = u_3^{RMS}L/\nu$ as a function of both Ra and Pr. The data in figure \ref{fig:figureRe}a correspond to the low $\tpr = 0.7$ and low $\Gamma = 0.5$ parameters, where we expect a large difference. This is confirmed for Nu in figure \ref{nura}b and appears to be the same for $\tre_3^{RMS}$. A similar difference in structure between 2D and 3D can be seen, with no noticeable convergence at the highest evaluated Ra. $\tre_3^{RMS}$ can be seen to locally scale larger than $\tra^{1/2}$ for 2D, which highlights the roll state dependence of integral quantities in 2D. The comparison of $\tre_3^{RMS}$ as a function of Pr in figure \ref{fig:figureRe}b reveals a similar picture as seen by \cite{sch04} for $\tra = 10^6$: The Reynolds number of 2D converges to the 3D value at high Pr. In both cases the 2D $\tre_3^{RMS}$ is higher than in 3D, while in contrast Nu is lower in 2D compared to 3D. The inverse energy cascade in 2D is possibly causing a stronger LSC than in 3D. However, up to now there have been no studies on the existence of the inverse energy cascade in 2D RB and therefore this remains uncertain. It could also be that in 2D, all emitted plumes drive the LSC while in 3D not all plumes follow the motion of the LSC. This can result in a lower Nu due to plumes being dragged down by the LSC before releasing most of their thermal energy at the boundary opposite to the plumes' origin. In this situation $\tre_3^{RMS}$ can be higher in 2D while Nu is lower. 

\section{Boundary layer profile}
\begin{figure}
\centering
\subfigure{\includegraphics[width=0.59\textwidth]{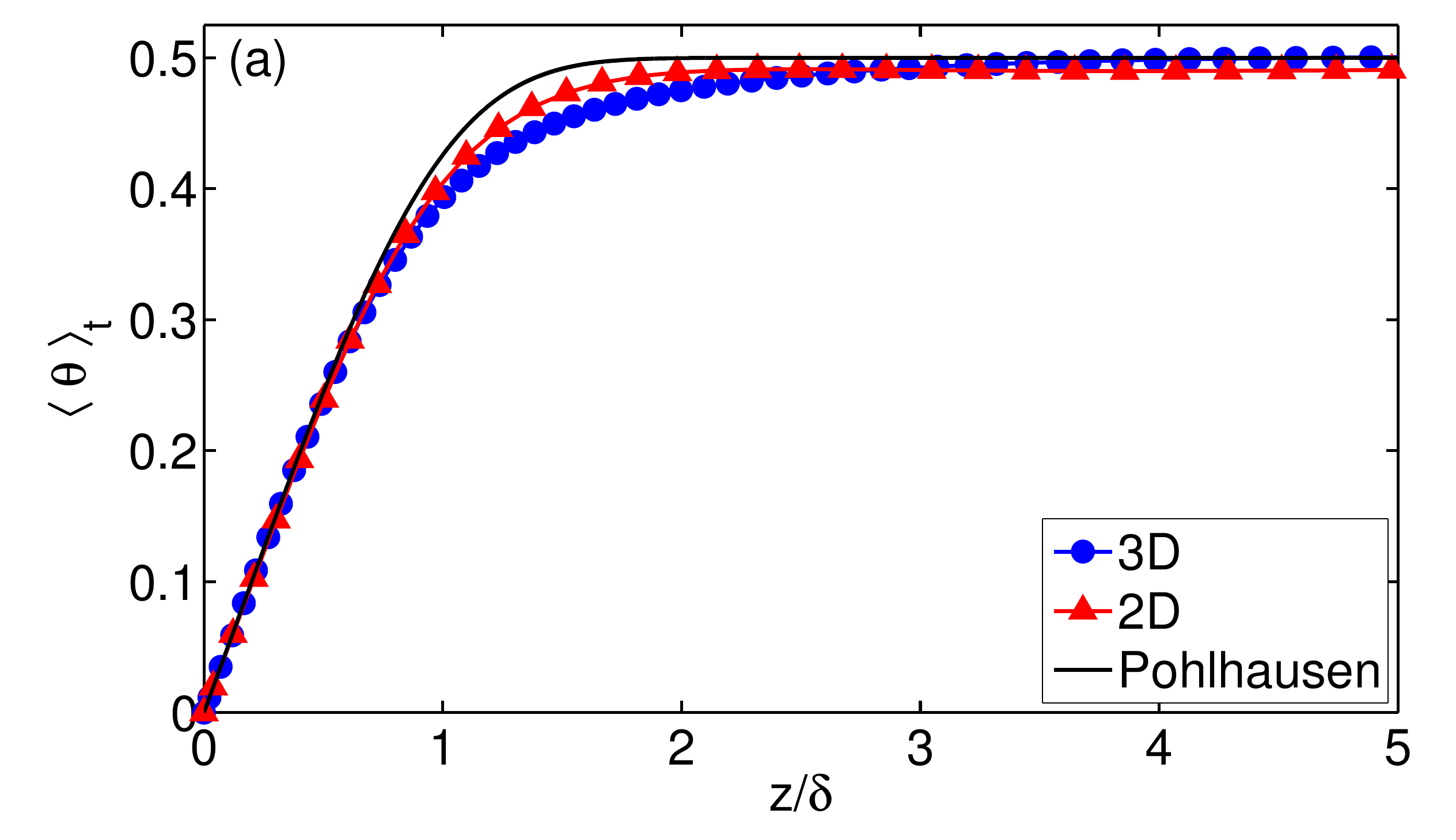}}
\subfigure{\includegraphics[width=0.40\textwidth]{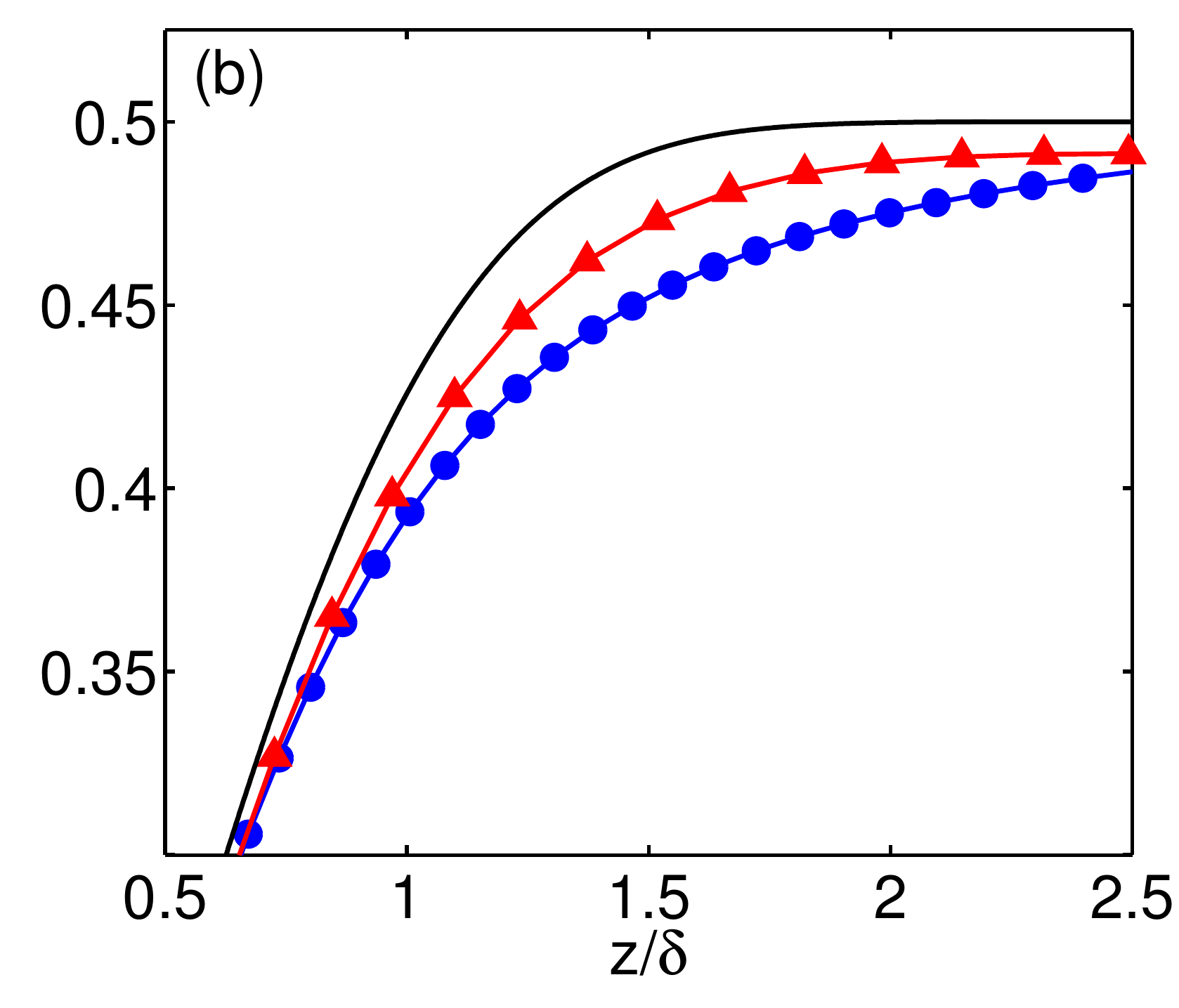}}
\caption{a) Thermal boundary layer profile for 2D (red triangles) and 3D (blue dots) compared to the Pohlhausen solution (black line) of the Prandlt-Blasius laminar boundary layer approximation. The 2D and 3D cases used for the profiles have identical $\tpr=4.38$ and Ra is varied to obtain an approximately equal Nu. For 3D $\tra = 10^8$ and $\tnu = 33.11$ and for 2D $\tra = 2.25 \cdot 10^8$ and $\tnu = 33.26$. b) A zoom of a).}
\label{fig:figure4}
\end{figure}

The boundary layer profile is a fundamental ingredient in most theoretical studies on the scaling of Nu and Re. In the 'classical' regime, where the \textbf{¥}boundary layer is assumed to be laminar, a reference analytical solution for the situation of a flow over a infinitely long plate is provided by \cite{poh21a}, which is based on the Prandtl-Blasius (PB) boundary layer approximation. The purpose of this section is to compare the 2D and 3D boundary layer profiles, with the Pohlhausen profile included for reference. For a laminar boundary layer, it is assumed that for $\tpr \approx 1$ both the velocity and temperature have a similar profile. This allows us to use the temperature boundary layer profile that is relatively easy to extract, for comparison. It is known that the time-averaged and instantaneous laminar boundary layers at the center of a large scale roll in both 2D and 3D RB flow are well approximated by the Pohlhausen profile when dynamically rescaled (\cite{zho10,zho10b,ste12}). This is despite the fact that the instantaneous flow in RB is only in rare cases locally parallel to the plates, in contrast with the PB assumptions of a completely parallel flow. The resulting deviations have been studied in detail for several cases \cite{wag12,shi12,sch12}. The vertical velocity gradient is non-zero due to the LSC, plume emission and corner rolls. As this effect is minimal at the center of a roll for most control parameters, the temperature profile is measured in the center of the cell; $r=0$ for 3D and $x=D/2$ for 2D in a $\Gamma = 1$ cell. It was shown by \cite{zho10b} that the lateral dependency is strong. In figure \ref{fig:figure4} the time-averaged temperature profiles for 2D and 3D for identical $\tpr = 4.38$ and the Pohlhausen solution are shown. The Ra number is varied to match Nu in 2D and 3D to obtain an equal temperature boundary layer thickness $\lambda_\theta\approx 1/(2\tnu)$ and similar local flow conditions induced by the heat flux. The profiles are measured in the lab frame in order to reveal the differences, as both the 2D and 3D profiles match the pohlhausen profile when measured in the dynamical frame.

It can be seen that both the 2D and 3D profiles in figure \ref{fig:figure4} do not match the Pohlhausen profile well in the labframe. The agreement of the 3D profile is worse than that of the 2D profile. This is most likely due to a combination of several causes. The PB theory is a 2D theory and due to the more complex dynamics of the LSC in 3D compared to 2D, the velocity field cannot be considered constantly parallel to the horizontal plates, even at $r=0$. Furthermore, increased plume activity in the bulk indicates that more plumes are emitted at the center of the 3D cell due the increased degrees of freedom and LSC cessations. This results in the 3D profile to be lower than 2D throughout the BL, as an increased amount of thermal energy is taken by plumes.

\section{Conclusion}
The comparability of two -and three-dimensional Rayleigh-B\'enard convection can in most cases be explained using the coherent structures present in the flow. At high Pr, the mushroom-type plume dominated regime, expected similar  2D and 3D behavior is observed. However, the LSC in 2D has a largely different effect on heat transport compared to 3D. In 2D the LSC covers the full system causing the plume movement to be dominated by the LSC, resulting in a significant discrepancy in this regime. 
The Ra and Pr scaling of the integral quantities in 2D and 3D are similar in some parameter regions. For $\tra = 10^8$, Nu appears similar for low and high Pr while it is substantially different for $\tpr \approx 1$. The similarity at low Pr is surprising as \cite{sch04} concluded, albeit for $\tra = 10^6$, that here 2D and 3D become incomparable. For $\tpr = 4.3$ the Nu(Ra) scaling is nearly identical with only a constant factor between them up to $\tra = 10^{10}$. The temperature boundary layer profiles of both 2D and 3D, obtained in the lab frame, differ from the Pohlhausen profile and from each other. As expected the 2D boundary layer is closer to the Pohlhausen profile. 

It is not difficult to find parameters for which there is a large difference between 2D and 3D. At low aspect-ratios the flow states in 2D vary more strongly than in 3D and have a larger effect on Nu and Re, in particular for $\tpr < 1$ (\cite{poe11}). This is reflected in the Nu(Ra) analysis at $\Gamma = 0.5$, where the 3D scaling is smooth and the 2D scaling is very structured. Less expected is the deviation at $\tpr = 4.3$ for $\tra > 10^{10}$, which concurs with a change in flow state in 2D. At this Ra the LSC breaks up and one would expect more similarity as the LSC in 3D differs strongly to the 2D LSC in that it does not limit the movement of plumes as much. Adding to the question is the discrepancy in Nu(Ra) around $\tpr = 1$. Here, the flow state is a LSC resulting in decreased similarity in contrast with the increased similarity in the Nu(Ra) scaling. 

A remarkable difference is found for $\tre_3^{RMS}$, which is in contrast to Nu, higher in 2D than in 3D. This can be attributed to the strong LSC in 2D, dragging thermal plumes back towards their originating plates before they can release their thermal energy.

\FloatBarrier
\begin{acknowledgments}
\noindent
{\it Acknowledgment:} The authors acknowledge useful discussions with Roberto Verzicco and Siegfried Grossmann. This study is supported by FOM and the National Computing Facilities (NCF), both sponsored by NWO. This work was granted access to the HPC resources of SARA made available within the Distributed European Computing Initiative by the PRACE-2IP, receiving funding from the
European Community's Seventh Framework Programme (FP7/2007-2013) under
grant agreement n¡ RI-283493. 

\end{acknowledgments}

\bibliographystyle{jfm}
\bibliography{2D3D_JFM.bbl}

\newpage
\appendix

\section{Details of numerical simulations}
Table \ref{table1} and \ref{table2} summarize the details of the 3D and 2D simulations that are presented in this study. The data are presented in a similar way as in table 1 of \cite{ste10}. The tables indicate the used grid resolution for the different Pr number cases and compare the resolution used in the boundary layer with the criterion given in equation (42) of \cite{shi10}. The resolution over the whole domain is compared by using equation (2.5) and (2.6) of \cite{ste10} and using $h=max(\Delta x,\Delta y, \Delta z)$ or $h=max(\Delta r, \Gamma L/2 \Delta \phi, \Delta z)$ in a cylindrical domain. Note that for high Pr number regime equation (2.6) of \cite{ste10} is more restrictive than the criterion given in equation (37) of \cite{shi10}. The current results seem to indicate that the criterion of \cite{shi10} is sufficient to assure convergence of the Nusselt number for the high Pr number cases.

In table \ref{table1} it can be seen that the number of gridpoints in the thermal boundary layer is not more than given by the criterion. However, the resolution tests show that the simulation is well resolved. In addition, the new parameter value $a = 0.911$ is used in the determination of this criterion, while numerical tests indicate that that the minimum number of gridpoints required in the BL is closer to the value obtained by using the old $a = 0.482$.

\begin{table}
\centering
\caption{Details for 3D simulations at $\tra=10^8$. The columns from left to right indicate the Pr number, the resolution in azimuthal, radial, and axial direction $N_\theta \times N_r \times N_z$, the number of points in the thermal boundary layer used in the simulation, the minimum number of points that should be used in the thermal boundary layer according to the criteria of \cite{shi10}, the average length scale in the flow compared to the largest grid length used somewhere in the grid $max(\delta_z,\delta_\phi,\delta_r)/\eta$, $\tnu_f$, i.e. the Nusselt number over the whole simulation length, without the initialization period that is disregarded, $\tnu_h$ the Nusselt number over the last half of the considered simulation time, $\tau_f$ the simulation time in free fall time units after the initialization period of $30$ to $200$ $\tau_f$ that is considered, and the reference where the simulation is first presented. Note that the bold cases indicate Pr number for which the effect of the numerical resolution has been tested. It can be seen in the table that the boundary layers are underresolved for low Pr according to the \cite{shi10} criterion. However, the agreement of Nu for identical parameters and different grid resolutions, indicate that at least for integral quantities the resolution is sufficient and that the criterion is possible too strict for low Pr.}
\FloatBarrier
    \begin{tabular}{ | c | c | c | c | c | c | c | c |c |}
    \hline
    Pr 				&$N_\theta \times N_r \times N_z$	&  $N_{BL}$	&  $N_{BL}^{min}$	& $\frac{max(\delta_z,\delta_\phi,\delta_r)}{\eta}$ &	  $\tnu_f$	 & $\tnu_h$& $\tau_f$ & Ref. \\
    \hline
    {\bf55.00 }			& ${\bf384\times 192 \times 384}$ 		&	{\bf15}		& {\bf 8}	& 	{\bf1.68} & {\bf32.00}	& {\bf32.00}	& {\bf600}	& \cite{ste10a}	\\
    55.00 & $256\times 128 \times 256$ 		&	23		& 8		& 	2.52 & 32.25	& 32.20	& 750	& \cite{ste10a}	\\
    30.00 			& $256\times 128 \times 256$ 		&	15		& 8		& 	2.17 & 32.31	& 32.40	& 600	& \cite{ste10a}	\\
    20.00 			& $256\times 128 \times 256$ 		&	15		& 8		&	1.97 & 32.54	& 32.56	& 400	& \cite{ste10a}	\\
    15.00 			& $256\times 128 \times 256$	 	&	15		& 8		&	1.83 & 32.36	& 32.60	& 400	& \cite{zho09b}\\
    10.00 			& $256\times 128 \times 256$ 		&	15		& 8		&	1.65 & 32.42	& 32.53	& 400	& \cite{zho09b}	\\
    {\bf 6.400}	& ${\bf 384\times 192 \times 384}$ 		&	{\bf 23}		& {\bf 8}		&	{\bf 0.99}	& {\bf 32.59}	& {\bf 32.42}	& {\bf 400}	& \cite{ste10a}	\\
        6.400 			& $256\times 128 \times 256$ 		&	15		& 8		&	1.48 & 32.95	& 33.00	& 200	& \cite{zho09b}	\\
    4.380 			& $256\times 128 \times 256$ 		&	15		& 8		&	1.35 & 33.15	& 32.91	& 400	& \cite{zho09b}	\\
    3.050 			& $256\times 128 \times 256$ 		&	15		& 8		&	1.24 & 33.48	& 33.45	& 400	& \cite{zho09b}	\\
    1.500 			& $256\times 128 \times 256$ 		&	15		& 8		&	1.03	& 33.13	& 33.13	& 400	& \cite{zho09b}	\\
    {\bf0.700 	}		& ${\bf512\times 128 \times 256}$ 		&   	{\bf12}		& {\bf9}		& 	{\bf0.58} & {\bf31.71}	& {\bf31.79}	& {\bf302}	& this work	\\
     0.700		 & $256\times 128 \times 256$ 		&	15		& 9		&	1.11	& 31.94	& 31.88	& 400	& \cite{zho09b}	\\
    0.450 			& $512\times 128 \times 256$ 		&	12		& 12	&	0.73 & 31.13	& 31.22	& 259	& this work	\\
    0.300 			& $512\times 128 \times 256$ 		&	13		& 15	&	0.88 & 30.00	& 30.03	& 341	& this work	\\
    {\bf0.200 }			& ${\bf1024\times 256 \times 512}$ 	&	{\bf27}		& {\bf18}	&	{\bf0.53}	& {\bf28.40}	& {\bf28.64}	& {\bf151}	& this work	\\
        0.200 & $512\times 128 \times 256$ 		&	13		& 18		& 	1.07 & 28.73	& 28.68	& 377	& this work	\\
    0.140 			& $1024\times 256 \times 512$ 	&	28		& 22		&	0.63 & 27.27	& 27.37	& 135	& this work	\\
    0.100 			& $1024\times 256 \times 512$ 	&	29		& 27		&	0.74 & 25.93	& 25.84	& 123	& this work	\\
    0.065 			& $1024\times 256 \times 512$ 	&	30		& 35		&	0.90	& 24.07	& 23.64	& 126	& this work	\\
    0.045 			& $1024\times 256 \times 512$ 	&	33		& 45		&	1.05	& 22.38	& 22.26	& 100	& this work	\\
    {\bf0.030}			& ${\bf1024\times 256 \times 512}$ 	&	{\bf39}		& {\bf57}	&	{\bf1.35} & {\bf20.31}	& {\bf20.04}	& {\bf56}		& this work	\\
        0.030  & $512\times 128 \times 256$ 		&	18		& 57		&	2.55 & 20.20	& 20.48	& 447	& this work	\\
     {\bf0.020}			 & ${\bf1024\times 256 \times 512}$ 	&	{\bf44}		& {\bf76}&	{\bf1.61}	& {\bf18.90}	& {\bf19.46}	& {\bf78}		& this work	\\
         0.020 & $512\times 128 \times 256$ 		&	19		& 76	&	3.04 & 18.82	& 18.59	& 414	& this work	\\
    \hline
  \end{tabular}  
\label{table1}
\end{table}

\begin{table}
\centering
\caption{Details for 2D simulations at $\tra=10^8$. The columns from left to right indicate the Pr number, the resolution in horizontal and vertical direction $N_x \times N_y$, the number of points in the thermal boundary layer used in the simulation, the minimum number of points that should be used in the thermal boundary layer according to the criteria of \cite{shi10}, the average length scale in the flow compared to the largest grid length used somewhere in the grid $max(\delta_x,\delta_y)/\eta$ and $\tau_f$ the simulation time in free fall time units after the initialization period of $30$ to $3000$ $\tau_f$ that is considered. Note that the bold cases indicate Pr number for which the effect of the numerical resolution has been tested.}
    \begin{tabular}{ | c | c | c | c | c | c | c | c |}
    \hline
    Pr 				&$N_x \times N_y$	&  $N_{BL}$	&  $N_{BL}^{min}$	& $\frac{max(\delta_x,\delta_y)}{\eta}$ &$\tnu_f$	 & $\tnu_h$& $\tau_f$\\
    \hline
   {\bf 55.00} 		& ${\bf 512\times 256 }$	&	{\bf 16}	& {\bf 7}	& 	{\bf 1.27}	& {\bf 27.83}	& {\bf 27.86}	& {\bf 50000}\\
    55.00 			& $256\times 128$ 		&	8		& 7		& 	2.38 		& 27.76 	& 27.75 	& 50000\\
    30.00 			& $256\times 128$ 		&	8		& 7		& 	2.38 		& 27.50 	& 27.51	& 40000\\
    20.00 			& $256\times 128$ 		&	8		& 7		&	2.38 		& 27.44 	& 27.44	& 30000\\
    10.00 			& $256\times 128$	 	&	8		& 7		&	2.37 		& 27.25 	& 27.26	& 20000\\
    {\bf 6.400}		& ${\bf 512\times 256}$ 		&	{\bf 16}	& {\bf 7}	&	{\bf 1.26}	& {\bf 26.99}	& {\bf 26.99}& {\bf10000}\\
    6.400 			& $256\times 128$ 		&	8		& 7		&	2.36 		& 26.79 	& 26.79	& 20000\\
    4.380 			& $256\times 128$ 		&	8		& 7		&	2.33 		& 25.51	& 25.47 	& 30000\\
    3.000 			& $256\times 128$ 		&	8		& 7     	&	2.32 		& 25.07	& 25.05	& 10000\\
    1.500 			& $256\times 128$ 		&	8		& 7		&	2.31		& 24.83	& 24.82	& 10000\\
    1.000 			& $512\times 256$ 		&   	8		& 7		& 	2.32 		& 25.17	& 25.16	& 10000\\
    {\bf 0.700} 		&  ${\bf1024\times 512}$		&	{\bf 30}		& {\bf 8}		&	{\bf0.78}		& {\bf25.22} 	& {\bf 25.10}	& {\bf  8000}\\
    0.700 			& $512\times 256$ 		&   	15		& 8		& 	1.48 		& 25.14	& 25.16	& 8000\\
    0.450 			& $1024\times 512$ 		&	34		& 11		&	0.96 		& 24.78	& 24.59 	& 2000\\
    0.300 			& $1024\times 512$ 		&	37		& 13		&	1.18 		& 25.25	& 25.09 	& 2000\\
    0.200 			& $1024\times 512$ 		&	34		& 17		&	1.43		& 25.73	& 25.40	& 1500\\
    {\bf 0.100} 		&  ${\bf2048\times 1024}$ 	&	{\bf 76}		& {\bf 26}	&	{\bf 1.01} 		& {\bf 23.63}	& {\bf 23.18}	& {\bf 200}\\
    0.100 			& $1024\times 512$ 		&	36		& 26		&	1.97 		& 23.32	& 23.03	& 1000\\
    0.065 			& $2048\times 1024$ 	&	82		& 34		&	1.23		& 21.76	& 22.07	& 200\\
    0.045 			& $2048\times 1024$ 	&	102		& 43		&	1.46		& 20.52	& 20.95	& 80\\
    \hline
  \end{tabular}  
\label{table2}
\end{table}

\end{document}